\documentstyle[psfig,12pt]{article}

\def\beq{\begin{equation}}
\def\eeq{\end{equation}}

\def\be{\begin{equation}}
\def\bea{\begin{eqnarray}}
\def\ee{\end{equation}}
\def\eea{\end{eqnarray}}
\def\d{\partial}

\begin{document}


\begin{titlepage}

\begin{flushright} UMN-D-02-5 \end{flushright}

\begin{centering}

\vspace*{3cm}

{\Large\bf Properties of the Bound States of Super-Yang--Mills-Chern--Simons
Theory}

\vspace*{1.5cm}

{\bf J.R.~Hiller,$^a$ S.S.~Pinsky,$^b$ and U.~Trittmann$^b$}
\vspace*{0.5cm}

$^a$
{\sl Department of Physics \\
University of Minnesota Duluth\\
Duluth, MN~~55812~~USA}

\vspace*{0.5cm}

$^b$
{\sl Department of Physics\\
Ohio State University\\
Columbus, OH~~43210~~USA}

\vspace*{1cm}

{22 August 2002}

\vspace*{2cm}


\vspace*{1cm}

\begin{abstract}
We apply supersymmetric discrete light-cone quantization (SDLCQ)  
to the study of supersymmetric Yang--Mills-Chern--Simons (SYM-CS) 
theory on ${\bf R} \times S^1 \times S^1$\@.  One of the compact 
directions is chosen to be light-like and the other to be space-like. 
Since the SDLCQ regularization explicitly preserves supersymmetry,  
this theory is totally finite, and thus we can solve for
bound-state wave functions and masses numerically without renormalizing.
The Chern--Simons term is introduced here to provide masses for the 
particles while remaining totally within a supersymmetric context. We 
examine the free, weak and strong-coupling spectrum. The transverse 
direction is discussed as a model for universal extra dimensions in the 
gauge sector. The wave functions are used to calculate the structure 
functions of the lowest mass states. We discuss the properties of
Kaluza--Klein states and focus on how they appear at strong
coupling.  We also discuss a set of anomalously light states which are 
reflections of the exact Bogomol'nyi--Prasad--Sommerfield
states of the underlying SYM theory.
\end{abstract}
\end{centering}

\vfill

\end{titlepage}
\newpage

\section{Introduction}

In 2+1 dimensions Chern--Simons (CS) theories are some of the most
interesting field theories.  For a review of
these theories see~\cite{dunne}. There is also considerable work on
SYM-CS theories. These theories have their own
remarkable properties. It has been shown that there is a finite anomaly
that shifts the CS coupling~\cite{lee}, and it has been conjectured by
Witten~\cite{witten2} that this theory spontaneously breaks supersymmetry
for some values of the CS coupling.

It is certainly useful to  numerically simulate these theories,
since they are
interesting from so many different points of view. The method  we
will use is SDLCQ (Supersymmetric Discrete Light-Cone Quantization).  This
is a numerical method that can be used to solve any theory with enough
supersymmetry to be finite.  The central point of this method is that
using DLCQ~\cite{pb85,bpp98} we can construct a finite dimensional
representation of the
superalgebra~\cite{sakai95}. From this representation of the superalgebra,
we construct a finite-dimensional Hamiltonian which we diagonalize
numerically. We repeat the process for larger and larger representations
and extrapolate the solution to the continuum.

We have already solved (2+1)-dimensional supersymmetric Yang--Mills (SYM)
theories by this method~\cite{hpt2001,hpt2001b} and also the dimensionally 
reduced SYM-CS theory~\cite{SYMCS1+1,BPS1+1}. Two-dimensional CS theory is very
QCD-like.  The states with more partons tend to be more massive, and many of
the low-mass states have a valence-like structure. The states have a dominant
component of the wave function with  a particular number and type of parton. 
We found that (1+1) and (2+1)-dimensional ${\cal N}=1$ SYM theories have 
an interesting set of massless Bogomol'nyi--Prasad--Sommerfield (BPS) 
bound states~\cite{hpt2001,hpt2001b}.  
These states are reflected in the (1+1)-dimensional CS theory  
as a set of states whose masses are approximately independent of 
the YM coupling $g$ and 
equal to the square of the sum of the CS masses of the partons in the bound
state~\cite{SYMCS1+1,BPS1+1}.  Massless BPS states are also present in the 
(2+1)-dimensional SYM theory, and we again see their reflection in the 
(2+1)-dimensional SYM-CS theory~\cite{BPS2+1}. They have
anomalously light masses, and we have already published some results for 
one of these states~\cite{BPS2+1}. We will report more results here.

String theory suggests that a fundamental theory of the world has 
more than 3+1 dimensions. It has therefore become interesting to consider 
how higher dimensional theories behave when viewed from a universe
with a smaller number of dimensions. 
In particular the momentum of a particle in the extra
dimension plays the role of an additional mass in lower dimensions. 
Excitations of this degree of freedom are called Kaluza--Klein 
(KK) particles.  Finding such KK particles could be considered a sign 
of extra dimensions. 
Since we are focusing here on (2+1) and (1+1)-dimensional theories, 
it is natural to think of the transverse dimension as the extra 
dimension and look at its effect in the two-dimensional world. 
We can view the spectrum of this model as a model of a
(3+1)-dimensional world that is part of a (4+1)-dimensional ``fundamental"
theory. From this point of view, the states with nonzero transverse momentum
are KK states, i.e.~states that have contributions to their mass from the 
transverse momentum in the ``extra'' dimension. This
is a useful model to have because it allows us to identify KK particles
that are a part of bound states and ask how they will look both in the
higher dimensional space and in the reduced dimensional space. This will be
one of the main themes when we analyze the bound states of this theory.

We investigate the various physical effects that we just discussed in
terms of the structure functions of the bound states. We find that the
anomalously light states, which are a reflection of 
the BPS states of the SYM theory, have a  unique structure function. This 
gives hope that such states might be disentangled from the rest of the 
spectrum. We discuss the structure functions of KK states both at weak 
and strong coupling. We also present the structure functions of 
the states that are mixtures of KK states and low-energy states.

In constructing the discrete approximation we drop the
longitudinal zero-momentum mode.  For some discussion of
dynamical and constrained zero modes, see the review~\cite{bpp98}
and previous work~\cite{alpt98}.
Inclusion of these modes would be ideal, but the techniques
required to include them in a numerical calculation
have proved to be difficult to develop, particularly because
of nonlinearities.   For DLCQ calculations that can be
compared with exact solutions, the exclusion of
zero modes does not affect the massive spectrum~\cite{bpp98}.
In scalar theories it has been known for some time that
constrained zero modes can give rise to dynamical symmetry
breaking~\cite{bpp98}, and work continues on the role of zero modes and near
zero modes in these theories~\cite{thorn}.

Dropping the zero modes in CS theory does result in the loss of many of 
the interesting aspects of CS theory, most notably the quantization of 
the CS coupling.  However, one particularly interesting property that 
will be preserved is the fact that the CS term simulates a mass for 
the theory. It is well known that supersymmetric abelian CS theory
is simply the theory of a free massive fermion and a free massive boson. 
In the non-abelian theory additional interactions are introduced, but 
we expect to also see
this  mass effect. This is particularly interesting here because  reduced
${\cal N}=1$ SYM theories are very stringy.  The low-mass states are
dominated by Fock states with many constituents, and as the size of the
superalgebraic representation is increased, states with lower masses and
more constituents appear~\cite{hpt2001,hpt2001b,alpt98,Lunin:2001im,%
Pinsky:2000rn,Hiller:2000nf,Haney:2000tk,Lunin:1999ib,Antonuccio:1999zu,%
Antonuccio:1998mq,Antonuccio:1998tm,Antonuccio:1998jg,Antonuccio:1998kz}.
The connection between string theory and supersymmetric gauge theory leads
one to expect this type of behavior; however, these gauge theories are not
very QCD-like.  Ultimately one might like to make contact with the
low-mass spectrum observed in nature.

The remainder of the paper is structured as follows.  In
Sec.~\ref{sec:SuperCS} we provide a summary of SYM-CS theory and our
numerical approximation; much of this is given in~\cite{SYMCS1+1} but 
is repeated here for completeness.  
In Sec.~\ref{sec:stage}, we set the stage by reviewing what we already know 
about the pieces of this theory, including the free discrete spectrum
of pure CS theory.  In Sec.~\ref{sec:FullTheory} we present our main
results.  A discussion of weak coupling, which includes a nice picture of 
the KK modes, is given in Sec.~\ref{weak}.  The coupling dependence of the theory 
and our results at strong coupling are discussed in Sec.~\ref{coupling}. 
Then in Sec.~\ref{structure} we define structure functions for the bound 
states and discuss in terms of them the various physical phenomena.
Section~\ref{sec:summary} contains a summary of our conclusions and some 
discussion of what could be done next.

\section{Supersymmetric Chern--Simons theory} \label{sec:SuperCS}

We consider ${\cal N}=1$ supersymmetric CS theory in 2+1 dimensions.
The Lagrangian of this theory is 
\begin{equation}
{\cal L}={\rm Tr}(-\frac{1}{4}{\cal L}_{\rm YM}+i{\cal L}_{\rm F}
+\frac{\kappa}{2}{\cal L}_{\rm CS}),\label{Lagrangian}
\end{equation}
where $\kappa$ is the CS coupling and
\begin{eqnarray}
{\cal L}_{\rm YM}&=&F_{\mu\nu}F^{\mu\nu}\,, \\
{\cal L}_{\rm F}&=&\bar{\Psi}\gamma_{\mu}D^{\mu}\Psi\,, \\
{\cal L}_{\rm CS}&=&\epsilon^{\mu\nu\lambda}\left(A_{\mu}
\partial_{\nu}A_{\lambda}+\frac{2i}{3}gA_\mu A_\nu A_\lambda \right)
+2\bar{\Psi}\Psi\,. \label{eq:CSLagrangian}
\end{eqnarray}
The two components of the spinor $\Psi=2^{-1/4}({\psi \atop \chi})$
are in the adjoint representation of $U(N_c)$ or $SU(N_c)$ and are the
chiral projections of the spinor $\Psi$, also defined by
\begin{equation}
\psi=\frac{1+\gamma^5}{2^{1/4}}\Psi\,,\qquad
\chi=\frac{1-\gamma^5}{2^{1/4}}\Psi\,.
\end{equation}
We will work in the
large-$N_c$ limit.  The field strength and the covariant derivative are
\begin{equation}
F_{\mu\nu}=\partial_{\mu}A_{\nu}-\partial_{\nu}A_{\mu}
              +ig[A_{\mu},A_{\nu}]\,, \quad \quad
D_{\mu}=\partial_{\mu}+ig[A_{\mu},\quad]\,.
\end{equation} 
The supersymmetric variations of the fields are
\begin{eqnarray}
\delta A_\mu&=&i\bar{\epsilon}\gamma_{\mu}\Psi\,,  \\
\delta\Psi&=&\frac{1}{4}i\epsilon^{\mu\nu\lambda}\gamma_\lambda F_{\mu\nu}
=\frac{1}{4}\Gamma^{\mu\nu}\epsilon F_{\mu\nu}\,,
\end{eqnarray}
where
\begin{equation}
\gamma^0=\sigma_2, \quad \gamma^1=i\sigma_1,\quad
\gamma^2=i\sigma_3, \quad \Gamma^{\mu\nu}\equiv
\frac{1}{2}\{\gamma^\mu,\gamma^\nu\}=
          i \epsilon^{\mu\nu\lambda} \gamma_{\lambda}\,.
\end{equation}
This leads to the supercurrent $Q^{(\mu)}$ in the usual manner via
\begin{equation}
\delta{\cal L}=\bar{\epsilon}\partial_{\mu}Q^{(\mu)}.
\label{LQ}
\end{equation}
Light-cone coordinates in 2+1 dimensions are $(x^+,x^-,x^\perp)$ where
$x^+=x_-$ is the light-cone time and $x^\perp=-x_\perp$. The totally
anti-symmetric tensor is defined by $\epsilon^{+-2}=-1$.
The variations of the three parts of the Lagrangian in
Eq.~(\ref{Lagrangian}) determine the (`chiral') components $Q^\pm$ of the
supercharge
via Eq.~(\ref{LQ}) to be
\begin{equation}
\int d^2x Q^{(+)}=\left( {Q^+ \atop Q^-}\right)
   = \frac{i}{2}\int d^2x\,
\Gamma^{\alpha\beta}\gamma^{+}\Psi F_{\alpha\beta}\,.
\end{equation}
Explicitly they are
\begin{eqnarray} \label{supercharges}
Q^-&=&-i 2^{3/4}\int d^2x\,
\psi\left(\partial^+ A^- -\partial^-A^+ + ig[A^+,A^-]\right)\,,
\nonumber \\
Q^+&=&-i 2^{5/4}\int d^2x\,
\psi\left(\partial^+ A^2 -\partial^2 A^+ + ig[A^+,A^\perp]\right)\,.
\end{eqnarray}
One can convince oneself by calculating the energy-momentum tensor
$T^{\mu\nu}$ that the supercharge fulfills the supersymmetry algebra
\begin{equation}
\{Q^\pm,Q^\pm\}=2\sqrt2 P^\pm\,, \qquad
\{Q^+,Q^-\}=-4P^\perp\,.
\end{equation}

In order to express the supercharge in terms of the physical degrees of
freedom, we have to use equations of motion, some of which are
constraint equations. The equations of motion for the gauge fields are
\begin{equation}
D_\nu F^{\nu\alpha}=-J^\alpha\,,
\end{equation}
where 
\begin{equation}
J^\alpha=\frac{\kappa}{2}\epsilon^{\alpha\nu\lambda}
F_{\nu\lambda}+2g\bar{\Psi}\gamma^\alpha\Psi\,.
\end{equation}
For $\alpha=+$ this is a constraint for $A^-$,
\begin{equation}
D_-A^-=-(D_2-\kappa)A^\perp-\frac{1}{D_-}(D_2-\kappa)\partial_2 A^+
             +2g\frac{1}{D_-}\bar{\Psi}\gamma^+\Psi\,.
\end{equation}
In light-cone gauge, $A^+=0$, this reduces to
\begin{equation}
D_-A^-=\frac{1}{D_-}[(\kappa-D_2)D_-A^\perp+2g\bar{\Psi}\gamma^+\Psi]\,.
\end{equation}
The equation of motion for the fermion is
\begin{equation}
\gamma^\mu D_\mu \Psi=-i\kappa\Psi\,.
\end{equation}
Expressing everything in terms of $\psi$ and $\chi$ leads to the
equations of motion
\begin{eqnarray}
\sqrt{2}D_+\psi&=&(D_2+\kappa)\chi\,,  \\
\sqrt{2}D_-\chi&=&(D_2-\kappa)\psi\,,
\end{eqnarray}
the second of which is a constraint equation.  The constraint equations
are used to eliminate $\chi$ and $A^-$.

At large $N_c$ the canonical \mbox{(anti-)}commutators for
the propagating fields $\phi\equiv A_\perp$ and $\psi$ are,
at equal light-cone time $x^+$,
\begin{equation}
\left[\phi_{ij}(x^-,x_\perp),\d_-\phi_{kl}(y^-,y_\perp)\right]=
\left\{\psi_{ij}(x^-,x_\perp),\psi_{kl}(y^-,y_\perp)\right\}=
\frac{1}{2}\delta(x^- -y^-)\delta(x_\perp -y_\perp)\delta_{il}\delta_{jk}\,.
\label{comm}
\end{equation}
The expansions of the field operators in terms of creation and
annihilation operators for the Fock basis are
\bea
\lefteqn{
\phi_{ij}(0,x^-,x_\perp) =} & & \nonumber \\
& &
\frac{1}{\sqrt{2\pi L}}\sum_{n^{\perp} = -\infty}^{\infty}
\int_0^\infty
         \frac{dk^+}{\sqrt{2k^+}}\left[
         a_{ij}(k^+,n^{\perp})e^{-{\rm i}k^+x^- +{\rm i}
\frac{2 \pi n^{\perp}}{L} x_\perp}+
         a^\dagger_{ji}(k^+,n^{\perp})e^{{\rm i}k^+x^- -
{\rm i}\frac{2 \pi n^{\perp}}{L}  x_\perp}\right]\,,
\nonumber\\
\lefteqn{
\psi_{ij}(0,x^-,x_\perp) =} & & \nonumber \\
& & \frac{1}{2\sqrt{\pi L}}\sum_{n^{\perp}=-\infty}^{\infty}\int_0^\infty
         dk^+\left[b_{ij}(k^+,n^{\perp})e^{-{\rm i}k^+x^- +
{\rm i}\frac{2 \pi n^{\perp}}{L} x_\perp}+
         b^\dagger_{ji}(k^+,n^\perp)e^{{\rm i}k^+x^- -{\rm i}
\frac{2 \pi n^{\perp}}{L} x_\perp}\right]\,.
\nonumber
\eea
>From the field \mbox{(anti-)}commutators one finds
\begin{equation}
\left[a_{ij}(p^+,n_\perp),a^\dagger_{lk}(q^+,m_\perp)\right]=
\left\{b_{ij}(p^+,n_\perp),b^\dagger_{lk}(q^+,m_\perp)\right\}=
\delta(p^+ -q^+)\delta_{n_\perp,m_\perp}\delta_{il}\delta_{jk}\,.
\end{equation}
Notice that the compactification in $x^\perp$ means that
the transverse momentum modes are summed over a discrete
set of values $2\pi n^\perp/L$.  In order to have a finite
matrix representation for the eigenvalue problem, we must
truncate these sums at some fixed integers $\pm T$.  The
value of $T$ defines a physical transverse cutoff
$\Lambda_\perp=2\pi T/L$; however, given this definition,
$T$ can be viewed as a measure of transverse resolution
at fixed $\Lambda_\perp$.

The supercharge $Q^+$ takes the following form:
\begin{equation}
Q^+={\rm i}2^{1/4}\sum_{|n^\perp|\le T}\int_0^\infty dk\sqrt{k}\left[
b_{ij}^\dagger(k,n^\perp) a_{ij}(k,n^\perp)-
a_{ij}^\dagger(k,n^\perp) b_{ij}(k,n^\perp)\right]\,.
\end{equation}
The supercharge $Q^-$ can be written as
\begin{equation}
\label{Qminus}
Q^- = g Q^-_{\rm SYM}(T) + Q_\perp (T) + i\kappa Q^-_{\rm CS}(T)\,,
\end{equation}
where
\begin{equation}
\label{Qperp}
Q^-_\perp(T)=\frac{2^{3/4}\pi {\rm i}}{L}\sum_{|n^\perp|\le T}\int_0^\infty dk
\frac{n^\perp}{\sqrt{k}}\left[
a_{ij}^\dagger(k,n^\perp) b_{ij}(k,n^\perp)-
b_{ij}^\dagger(k,n^\perp) a_{ij}(k,n^\perp)\right]\,,
\end{equation}
\begin{equation}
\label{QCS}
Q^-_{\rm CS}(T)=2^{-1/4}{\rm i}\sum_{|n^\perp|\le T}\int_0^\infty dk
\frac{1}{\sqrt{k}}\left[
a_{ij}^\dagger(k,n^\perp) b_{ij}(k,n^\perp)+b_{ij}^\dagger(k,n^\perp)
a_{ij}(k,n^\perp)\right]\,,
\end{equation}
and
\begin{eqnarray}
\label{QSYM}
&&Q^-_{\rm SYM}(T)=
{{\rm i} 2^{-5/4} \over \sqrt{L\pi}}\sum_{|n_i^\perp|\le T}\int_0^\infty 
dk_1dk_2dk_3
\delta(k_1+k_2-k_3) \delta_{n^\perp_1+n^\perp_2,n^\perp_3}
\left\{2\left({ 1\over k_1}+{1 \over k_2}-{1\over k_3}\right)\right.\nonumber\\
&&\qquad \qquad \times\left[b_{ik}^\dagger(k_1,n^\perp_1) 
b_{kj}^\dagger(k_2,n^\perp_2)
b_{ij}(k_3,n^\perp_3)
+b_{ij}^\dagger(k_3,n^\perp_3) b_{ik}(k_1,n^\perp_1) 
      b_{kj}(k_2,n^\perp_2)]\right]        \nonumber \\
    && +{k_2-k_1  \over  k_3\sqrt{k_1 k_2}} 
\left[a_{ik}^\dagger(k_1,n^\perp_1) a_{kj}^\dagger(k_2,n^\perp_2)
b_{ij}(k_3,n^\perp_3)
-b_{ij}^\dagger(k_3,n^\perp_3)a_{ik}(k_1,n^\perp_1)
a_{kj}(k_2,n^\perp_2) \right]\nonumber\\
&&+{k_1+k_3  \over k_2 \sqrt{k_1 k_3}}
\left[a_{ik}^\dagger(k_3,n^\perp_3) a_{kj}(k_1,n^\perp_1) b_{ij}(k_2,n^\perp_2)
-a_{ik}^\dagger(k_1,n^\perp_1) b_{kj}^\dagger(k_2,n^\perp_2)
a_{ij}(k_3,n^\perp_3) \right]\nonumber\\
&&+{k_2+k_3  \over k_1\sqrt{k_2 k_3}}
\left[b_{ik}^\dagger(k_1,n^\perp_1) a_{kj}^\dagger(k_2,n^\perp_2)
a_{ij}(k_3,n^\perp_3)
-a_{ij}^\dagger(k_3,n^\perp_3)b_{ik}(k_1,n^\perp_1) 
      a_{kj}(k_2,n^\perp_2) \right] \left. \frac{}{}\right\}\,. \nonumber\\
\end{eqnarray}
For a massive particle the light-cone energy is $(k^2_\perp + m^2)/k^+$,
so $k_\perp$ behaves like a mass.  In comparing $Q_\perp^-$ and
$\kappa Q_{\rm CS}^-$ we see that $\kappa$ appears in a
very similar way to $k_\perp$.  Therefore $\kappa$ behaves in 
many ways like a mass.

To fully discretize the theory we impose periodic boundary
conditions on the boson and fermion fields alike, and obtain an expansion of
the fields with discrete momentum modes.
The discrete longitudinal momenta $k^+$ are written as fractions $nP^+/K$ of the
total longitudinal momentum $P^+$, with $n$ and $K$ positive integers. 
The longitudinal resolution is set by $K$, which is known in DLCQ
as the harmonic resolution~\cite{pb85}.
This converts the mass eigenvalue problem $2P^+P^-|M\rangle = M^2 |M\rangle$
to a matrix eigenvalue problem.
As discussed in the introduction, this is done
in SDLCQ by first discretizing the supercharge $Q^-$
and then constructing $P^-$ from the square of the supercharge:
$P^- = (Q^-)^2/\sqrt{2}$.
When comparing the $\perp$ and CS contributions to the supercharge, we
find a relative $i$ between them. Thus the usual eigenvalue
problem
\begin{equation}
2P^+P^-|M\rangle=
    \sqrt{2}P^+\left(g Q^-_{\rm SYM}(T) 
             + Q_\perp (T) + i\kappa Q^-_{\rm CS}(T)\right)^2|M\rangle
                  =M^2|M\rangle
\label{EVP}
\end{equation}
has to be solved by using fully complex methods.
The continuum limit is obtained by taking $T\rightarrow\infty$ and
$K\rightarrow\infty$.  The largest matrices are diagonalized by
the Lanczos technique~\cite{Lanczos} which easily yields several of the lowest
eigenvalues and their eigenvectors.  For a more complete discussion of
our numerical methods, see~\cite{BPS2+1} and \cite{hpt2001}.

We retain\footnote{We note that the CS term breaks transverse parity.}
the $S$-symmetry, which is associated with the orientation of the
large-$N_c$ string of partons in a state~\cite{kutasov93}. In a
(1+1)-dimensional model this orientation parity is usually referred as a
$Z_2$ symmetry, and we will follow that here.  It gives a sign when the color
indices are permuted
\begin{equation}\label{Z2}
Z_2 : a_{ij}(k)\rightarrow -a_{ji}(k)\,, \qquad
      b_{ij}(k)\rightarrow -b_{ji}(k)\,.
\end{equation}
We will use this symmetry to reduce the Hamiltonian matrix size and
hence the numerical effort. All of our states will be labeled by the
$Z_2$ sector in which they appear. We will not attempt to label the states
by their normal parity; on the light cone this is only an approximate
symmetry. Such a labeling could be done in an approximate way,
as was shown by Hornbostel~\cite{horn}, and might be useful for
comparison purposes if at some point there are results from
lattice simulations of the present theory.
 

\section{Limiting Cases}
\label{sec:stage}

Before considering results for the full theory we discuss what we have
learned about various limiting cases.  These are dimensionally reduced
pure SYM theory ($\kappa=0$, $T=0$)~\cite{Antonuccio:1998jg,Antonuccio:1998kz},
(2+1)-dimensional pure SYM theory ($\kappa=0$)~\cite{hpt2001},
dimensionally reduced SYM-CS theory ($T=0$)~\cite{SYMCS1+1,BPS1+1},
and pure CS theory ($g=0$).  These will provide convenient points
of reference when we discuss the full theory.

\begin{figure}
\begin{tabular}{cc}
\hspace*{-0.8cm}\psfig{figure=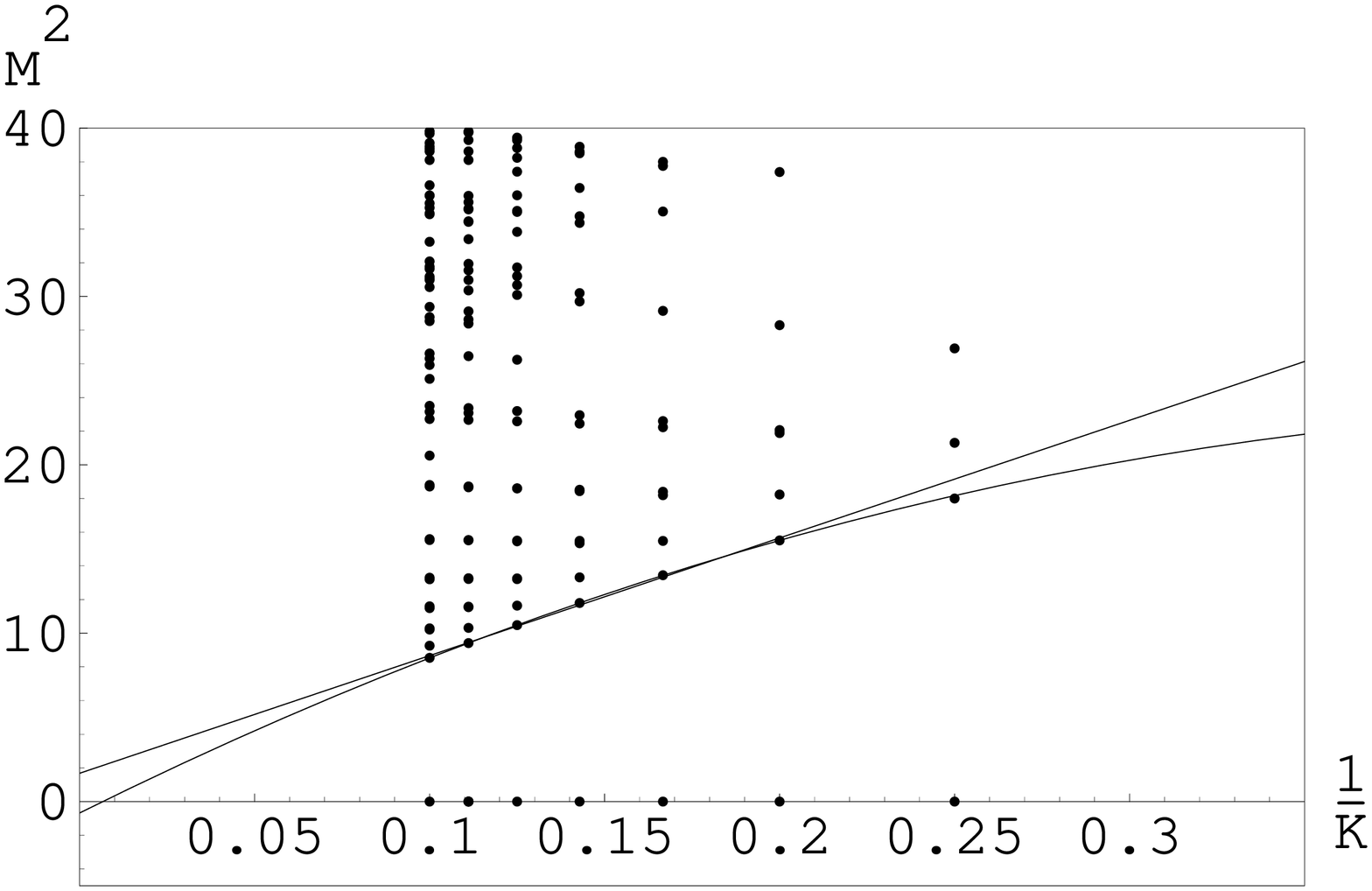,width=9.5cm,angle=-90}&
\hspace*{-1cm}\psfig{figure=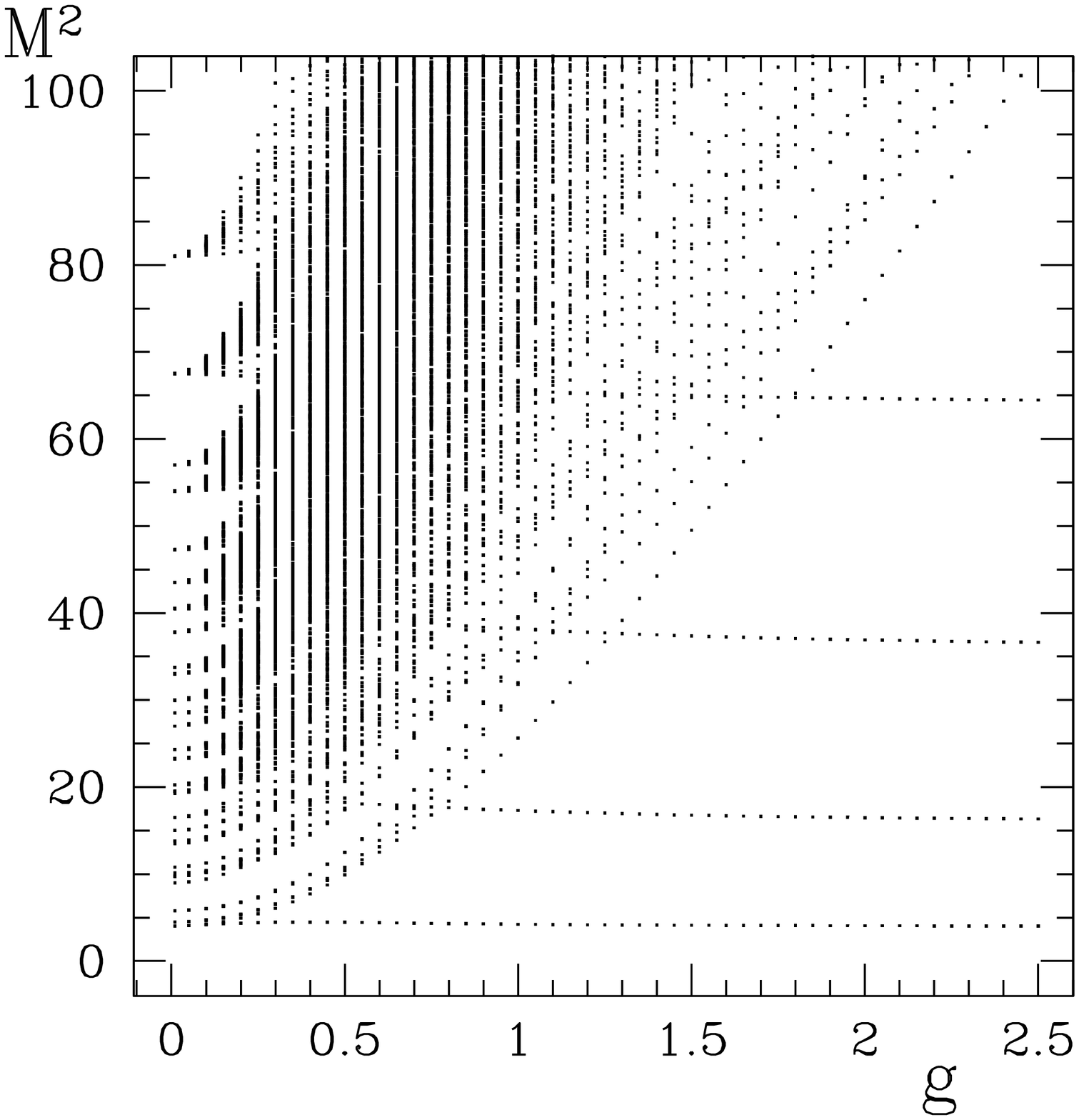,width=7.5cm,angle=0}\\
(a) & (b)
\end{tabular}
\caption{Spectrum of (a) (1+1)-dimensional SYM theory in units of $g^2 N_c/\pi$ and
(b) (1+1)-dimensional SYM-CS theory.  The masses are in
units of $\kappa^2$ at resolution $K=5$ 
\label{OLD} } 
\end{figure}

Let us start with dimensionally reduced
SYM theory~\cite{Antonuccio:1998jg,Antonuccio:1998kz}, for which
\begin{equation}
Q^-=gQ^-_{\rm SYM}(0)\,.
\end{equation}
There are two properties that characterize this theory. As we
increase the resolution we find that there are new lower mass states that
appear, and this sequence of states appears to accumulate at $M^2=0$. In
addition, there are massless BPS states, and the dominant component of the
wave function of this sequence of states can be arranged to have 2
particles, 3 particles, etc., up to the maximum number of particles allowed
by the resolution, {\em i.e.}, $K$ particles. Therefore at resolution $K$ 
there are $K-1$ boson BPS states and $K-1$ fermion BPS states. In
Fig.~\ref{OLD}(a) we see the spectrum of two-dimensional 
SYM as a function of the inverse of the resolution $1/K$ from
earlier work~\cite{Antonuccio:1998jg,Antonuccio:1998kz}. Also shown are 
two fits to the lowest mass state at each resolution which show that the 
accumulation point is consistent with zero.

When we extend the SYM theory to $2+1$ dimensions~\cite{hpt2001},
the supercharge $Q^-$ becomes
\begin{equation}
Q^-=gQ^-_{\rm SYM}(T) + Q^-_\perp(T)\,.
\end{equation}
Again there are a number of unique properties that characterize this 
theory.  At small coupling $g$  and low energy we recover completely 
the (1+1)-dimensional theory at $g=0$. At higher energies we find a 
series of KK states. This is an expected result and
was fully verified in earlier work. At large
$g$ one might expect that the $n_\perp/g$ term would freeze out 
and one would again see a reflection of the 1+1 theory. The fact 
is that while we see these states,
they are just a small part of the spectrum. $Q^-_{\rm SYM}$  
by itself wants to have a large number of particles, as we discussed 
above. The contributions from $gQ^-_{\rm SYM}$ and $Q^-_\perp$ are 
therefore minimized by a larger number of particles, each with a 
small transverse momentum.  We therefore find that the
average number of particles in the bound states grows with the YM coupling.
The massless BPS states of the 1+1 SYM persist here in 2+1 dimensions, but
now the number of particles in all of these states increases rapidly with 
YM coupling. 
 
When we add a CS term to the (1+1)-dimensional theory~\cite{SYMCS1+1},
we find that the most important
role of the CS term is to provide a mass for the constituents. This freezes
out the long, lower mass states that characterized (1+1)-dimensional SYM
theory.  Interestingly, however, the massless BPS states become massive
approximate BPS states and have masses that are nearly independent of 
the YM coupling~\cite{BPS1+1}, as seen in Fig.~\ref{OLD}(b).

If we set $g=0$ in Eq.~(\ref{Qminus}) we are left with the CS term and the
transverse momentum term. It is not surprising that the square of this
supercharge gives the light-cone energy of a set of noninteracting
massive adjoint fermions and
adjoint bosons. Understanding the discrete version of the continuum of
free-particle states in the DLCQ approximation is helpful, because once the
coupling is turned on SYM-CS theory becomes a confining theory and each of the
continuum states becomes a bound state. Therefore at small coupling we
expect to see bound-state masses close to the masses seen in the discrete
version of the continuum.

The discrete version of the two particle continuum spectrum is described by
\begin{equation} 
\frac{M^2(K)}{K}=\frac{\kappa^2 +k^2_{1,\perp}}{n_1}+
\frac{\kappa^2+k^2_{2,\perp}}{n_2}\,,
\label{cont2}
\end{equation} 
where $n_1+n_2=K$ and $k_{1\perp}+k_{2\perp}=0$. The lowest states in the
$Z_2=+1$ sector are part of the two-particle continuum spectrum.
{\scriptsize
\begin{table}
\centerline{
\begin{tabular}{|r|c||r|c|}
\hline
\multicolumn{4}{|c|}{$Z_2=1$}  \\
\hline
Fock state &$K{=}9$ &Fock state & $K{=}8$\\
\hline
$a(5,0)a(4,0)$ & 4.05& $a(4,0)a(4,0)$ & 4.00 \\
$b(5,0)b(4,0)$ & 4.05& $a(5,0)a(3,0)$ & 4.26 \\
$a(6,0)a(3,0)$ & 4.50& $b(5,0)b(3,0)$ & 4.26 \\
$b(6,0)b(3,0)$ & 4.50& $a(6,0)a(2,0)$ & 5.33 \\
$a(7,0)a(2,0)$ & 5.78& $b(6,0)b(2,0)$ & 5.33\\
$b(7,0)b(2,0)$ & 5.78& $a(4,-1)a(4,1)$ & 8.00\\
$a(5,-1)a(4,1)$ & 8.10& $b(4,-1)b(4,1)$ & 8.00\\
$b(5,-1)b(4,1)$ & 8.10& $a(5,-1)a(3,1)$ & 8.53\\
$a(5,1)a(4,-1)$ & 8.10& $b(5,-1)a(3,1)$ & 8.53\\
$b(5,1)b(4,-1)$ & 8.10& $a(5,1)a(3,-1)$ & 8.53\\
\hline
\end{tabular}
\begin{tabular}{|r|c||r|c|}
\hline
\multicolumn{4}{|c|}{$Z_2=-1$} \\
\hline
Fock state &$K{=}9$&Fock state &$K{=}8$\\\hline
$a(3,0)a(3,0)a(3,0)$&9.00&$a(3,0)a(3,0)a(2,0)$&9.33 \\
$a(2,0)b(3,0)b(4,0)$&9.75&$a(3,0)b(2,0)b(3,0)$&9.33\\
$a(4,0)a(3,0)a(2,0)$&9.75&$a(2,0)b(2,0)b(4,0)$&10.00\\
$a(4,0)b(2,0)b(3,0)$&9.75&$a(4,0)a(2,0)a(2,0)$&10.00\\
$a(3,0)b(2,0)b(4,0)$&9.75& $a(4,0)a(1,0)a(3,0)$&12.66\\
$a(2,0)b(2,0)b(5,0)$&10.80&$a(4,0)b(1,0)b(3,0)$&12.66\\
$a(5,0)a(2,0)a(2,0)$&10.80&$a(3,0)b(1,0)b(4,0)$&12.66\\
$a(4,0)a(4,0)a(1,0)$&13.50&$a(1,0)b(3,0)b(4,0)$&12.66\\
$a(4,0)b(1,0)b(4,0)$&13.50&$a(5,0)a(1,0)a(2,0)$&13.60\\
$a(5,0)a(3,0)a(1,0)$&13.80&$a(5,0)b(1,0)b(2,0)$&13.60\\
\hline
\end{tabular}
}
\caption{The Fock-state composition and mass squared 
$M^2$ in units of $4\pi^2/L^2$ for the continuum states in
the two-particle ($Z_2=1$) and three-particle ($Z_2=-1$) sectors 
for resolutions $K=9,8$.  The Chern--Simons coupling is
$\kappa=2\pi/L$; the Yang--Mills coupling is zero.
\label{table}
 }
\end{table}
}
The discrete version of the
three-particle continuum spectrum is given by
\begin{equation}
\frac{M^2(K)}{K}=\frac{\kappa^2 +k^2_{1,\perp}}{n_1}
    +\frac{\kappa^2 +k^2_{2,\perp}}{n_2}
     +\frac{\kappa^2 +k^2_{3,\perp}}{n_3}\,,
\label{cont3}
\end{equation}
where $n_1+n_2+n_3=K$ and $k_{1\perp}+k_{2\perp}+k_{3\perp}=0$.
The lowest states in the $Z_2=-1$ sector are part of the three-particle 
continuum spectrum.  The lowest mass states in the
$Z_2=+1$ and $Z_2=-1$ continuum spectrum, and their Fock-state composition,
are shown in Table~\ref{table}. Higher contributions are seen at the
appropriate places in the free particle spectrum as well. For example, 
among the higher mass states in the $Z_2=+1$ sector is the state
$a(3,0)b(3,0)b(3,0)$ which is degenerate with the lowest mass state 
in the $Z_2=-1$ sector.

In SYM theory in two and three dimensions the bound-state spectrum has a 
four-fold degeneracy with two boson states and two fermion states at each 
mass. When we add a CS term the theory becomes complex, and the degeneracy 
is reduced to a two-fold degeneracy, that is one boson and one fermion 
bound state at each mass.   At the special point $g=0$ we see in the 
table a number of accidental degeneracies, at least from the
supersymmetry point of view. In particular, the  two-parton states in 
the bosonic sector will consist of either two bosons or two fermions. 
We thus expect a double degeneracy of the multi-particle states, unless 
the fermion-fermion state is forbidden by the Pauli principle, 
as happens when the momenta are equal.

\section{Results for the Full Theory} \label{sec:FullTheory}

 \subsection{Weak coupling} \label{weak}

It is interesting to first consider weak coupling because it provides 
a clear connection between a variety of interesting phenomena that we 
want to consider in this paper. At low energy and small coupling the 
spectrum is essentially a (1+1)-dimensional spectrum. The lowest energy 
bound states are the two-parton states that we saw in the
$g=0$ spectrum, with a binding energy proportional to the YM coupling 
that breaks the accidental degeneracy at $g=0$. Since none of 
the partons have even one unit of transverse momentum at very low 
energy, the transverse momentum term in the
supercharge does not contribute, and the supercharge is the 
(1+1)-dimensional supercharge scaled by
$L$. It is therefore obvious that at low energy and small coupling
the (2+1)-dimensional spectrum is a reflection of the two-dimensional spectrum.

As we move up in energy at weak coupling we find the KK states. The 
masses are closely related to the masses of the KK states that we found 
at $g=0$. The lowest energy states are two-parton states with the partons 
having +1 and $-1$ units of transverse momentum. This essentially doubles 
the free energy of the bound state from 4 to 8. In addition, the transverse
momentum gives an interaction energy that
breaks the accidental degeneracies we saw at $g=0$.
\begin{figure}
\begin{tabular}{cc}
\psfig{figure=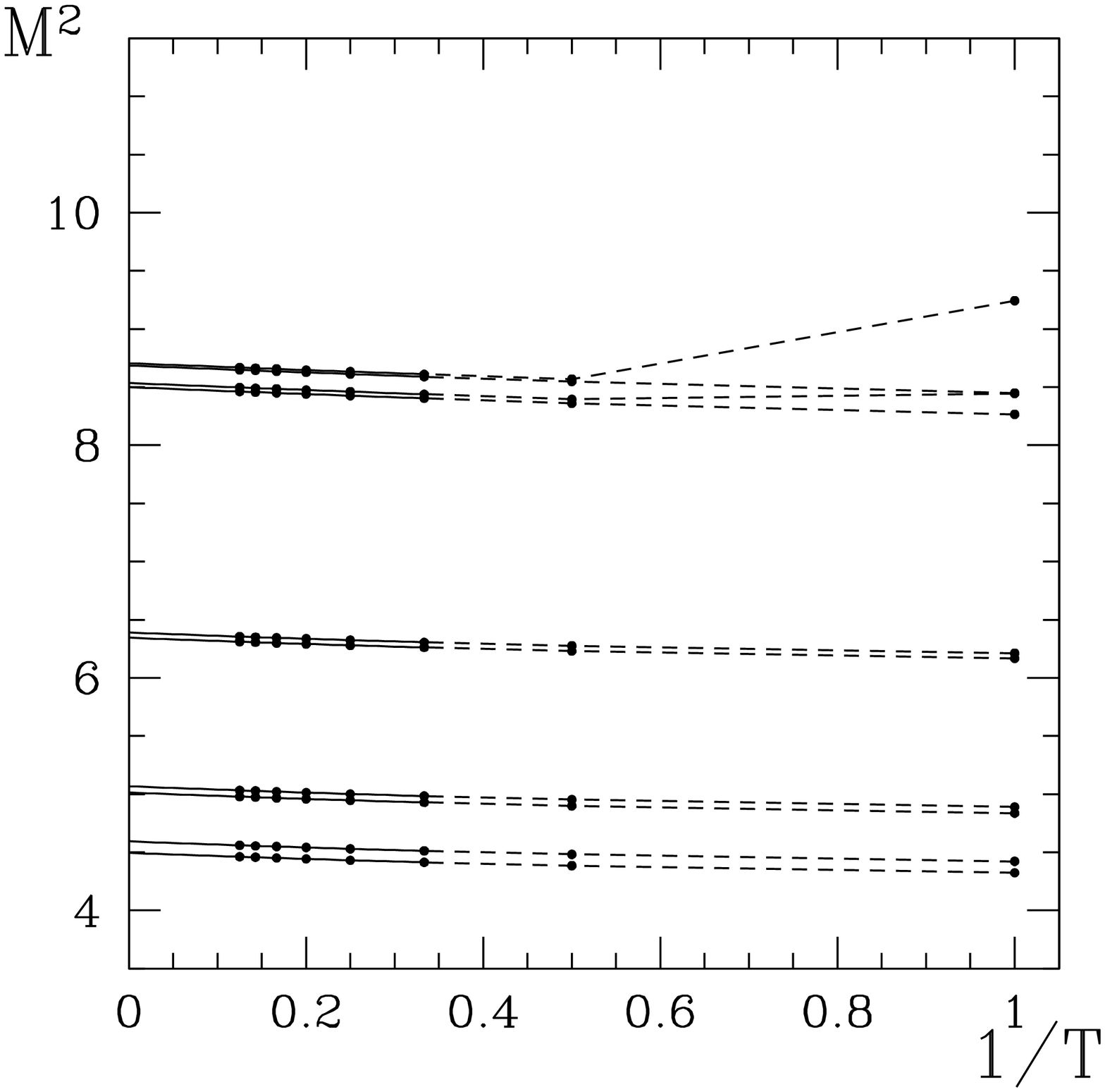,width=7.5cm,angle=0}&
\psfig{figure=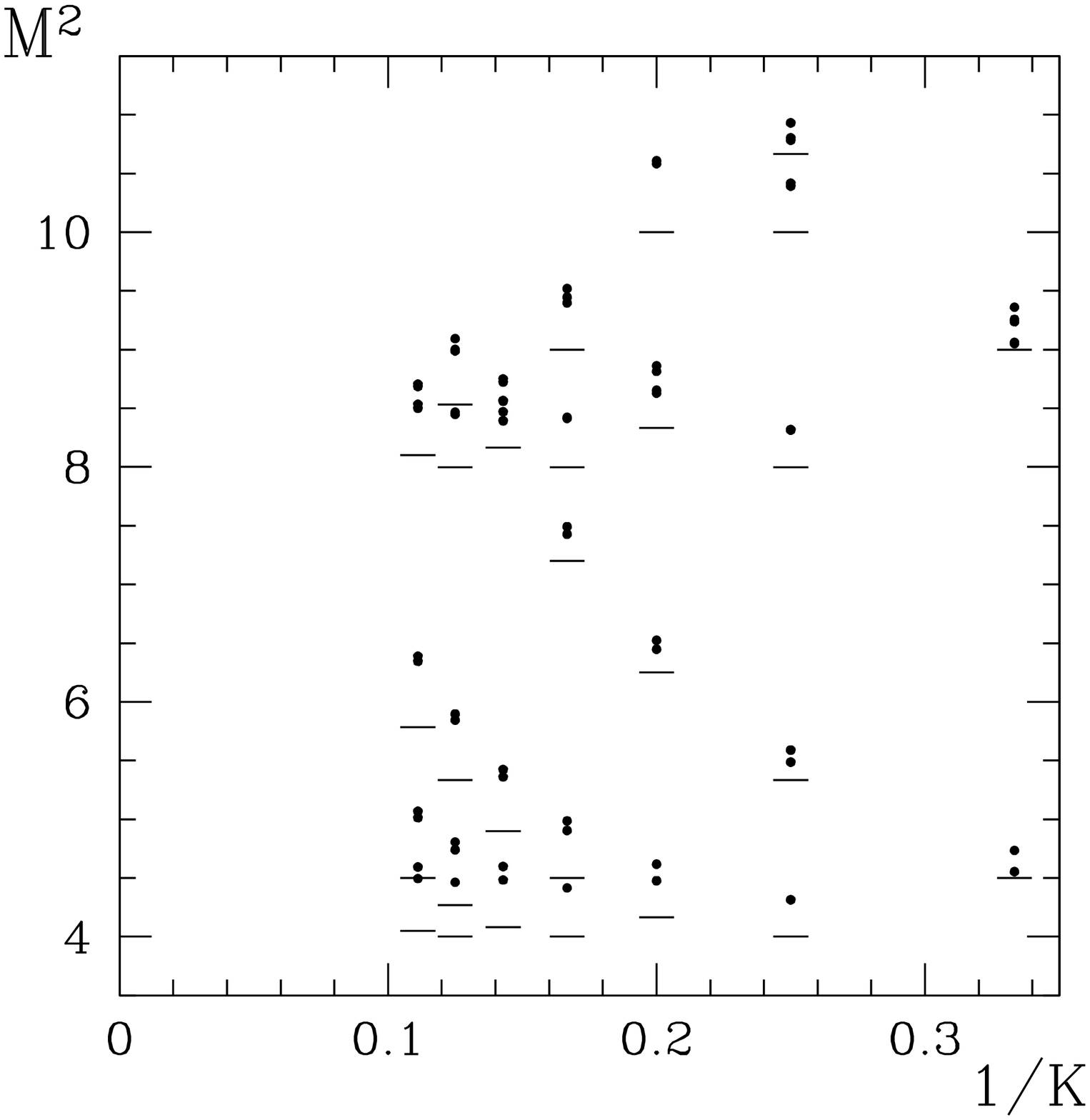,width=7.5cm,angle=0}\\
(a)&(b)
\end{tabular}
\caption{The spectrum as a function of (a) $1/T$, at $K=9$, and
(b) the longitudinal resolution $K$, for $g=0.1\sqrt{4\pi^3/N_cL}$.  
In (b) the horizontal lines
indicate the discrete continuum spectrum.\label{g01}
}
\end{figure}

We have done a detailed calculation for weak coupling 
with\footnote{For the remainder of the paper, values of $g$ will
be quoted in units of $\sqrt{4\pi^3/N_cL}$.} $g=0.1$. We 
gain an additional numerical advantage at small coupling in that the 
average number of particles is well behaved and well understood because 
each of the states is close to a continuum state. We are therefore able 
to put an additional cutoff on the number of partons in each state. For 
the ten lowest mass states in each sector we are able to limit ourselves 
to 4 partons. This allows us to carry the calculation to
9 units of transverse resolution and 9 units of longitudinal resolution. 
We have checked that calculating with and without this cutoff produces 
the same results at the highest values of the resolution where we can 
perform the complete calculation. 

We carry out a standard DLCQ analysis of the bound states. At each 
value of the resolution $K$ and in each $Z_2$ sector we look at the 
ten lowest energy bound states at each transverse resolution $T$.  
For $Z_2=+1$ we fit the curves of mass squared vs.\ $1/T$, as shown 
in Fig.~\ref{g01}(a), and extrapolate to infinite transverse momentum
cutoff. The intercepts of these curves are the infinite momentum
extrapolations, and we plot these as a function of $1/K$ in 
Fig.~\ref{g01}(b). The $Z_2=-1$ sector behaves very
similarly to the $Z_2=+1$ sector except that the lowest energy 
bound states consist primarily of three partons 
rather than two.

We plot the discrete version of the continuum spectrum as horizontal
lines at each resolution $K$ and the bound states as dots. We see that 
the accidental degeneracy we found at $g=0$ is completely broken, as 
expected. If we connect the lowest mass
states at each K we will get a saw tooth pattern. This saw tooth pattern 
is a reflection of the close connection of these states with the 
discrete continuum spectrum. We can nevertheless fit these points using
$M^2=M^2_\infty+b\frac{1}{K}$. The intercept of this curve yields the continuum
mass $M_\infty$ of the particular state.  We find for example that the lowest state 
is at approximately\footnote{Masses will be given in units of $4\pi^2/L^2$.}
$M^2=4.5$.

We note that at about $9.0$ we see the first KK excitation of these states.
In the section below on structure functions we will show the 
structure functions of these states and see the characteristic 
behavior of each of these types of states.

\subsection{Strong coupling} \label{coupling}
\begin{figure}
\centerline{
\psfig{figure=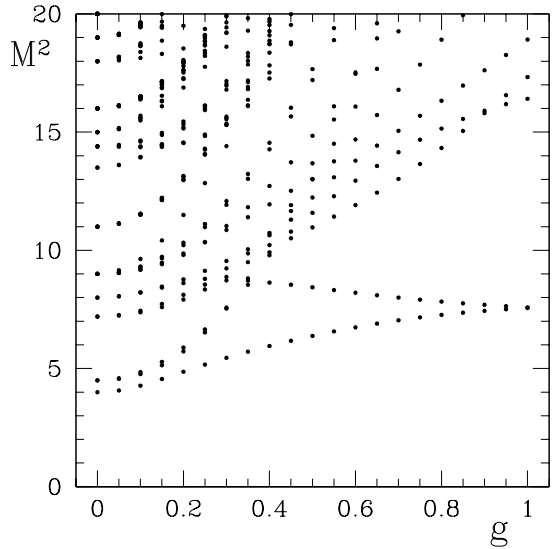,width=7.5cm,angle=0}}
\caption{The mass spectrum as a function of $g$, in units of 
$\sqrt{4\pi^3/N_cL}$, at $K=6$, $T=1$, and $Z_2=+1$.
\label{Mvsg}
}
\end{figure}

In Fig.~\ref{Mvsg} we show the coupling dependence of SYM-CS
theory in 2+1 dimensions. For numerical
convenience the results presented here are at specific and low
values of the longitudinal and transverse cutoffs $K$ and $T$ but 
at many values of the coupling. We can distinguish two regions for 
the low-lying spectrum.  For very small $g$ we have a spectrum of 
almost free particles as we discussed in the previous section. 
There is a transition that occurs around $g=0.3$ where we see a
number of level crossings, and beyond this point the character of
the spectrum changes. For larger $g> 0.3$ we will have a 
{\em bona fide} strong-coupling bound-state spectrum.

\begin{figure}
\begin{tabular}{cc}
\psfig{figure=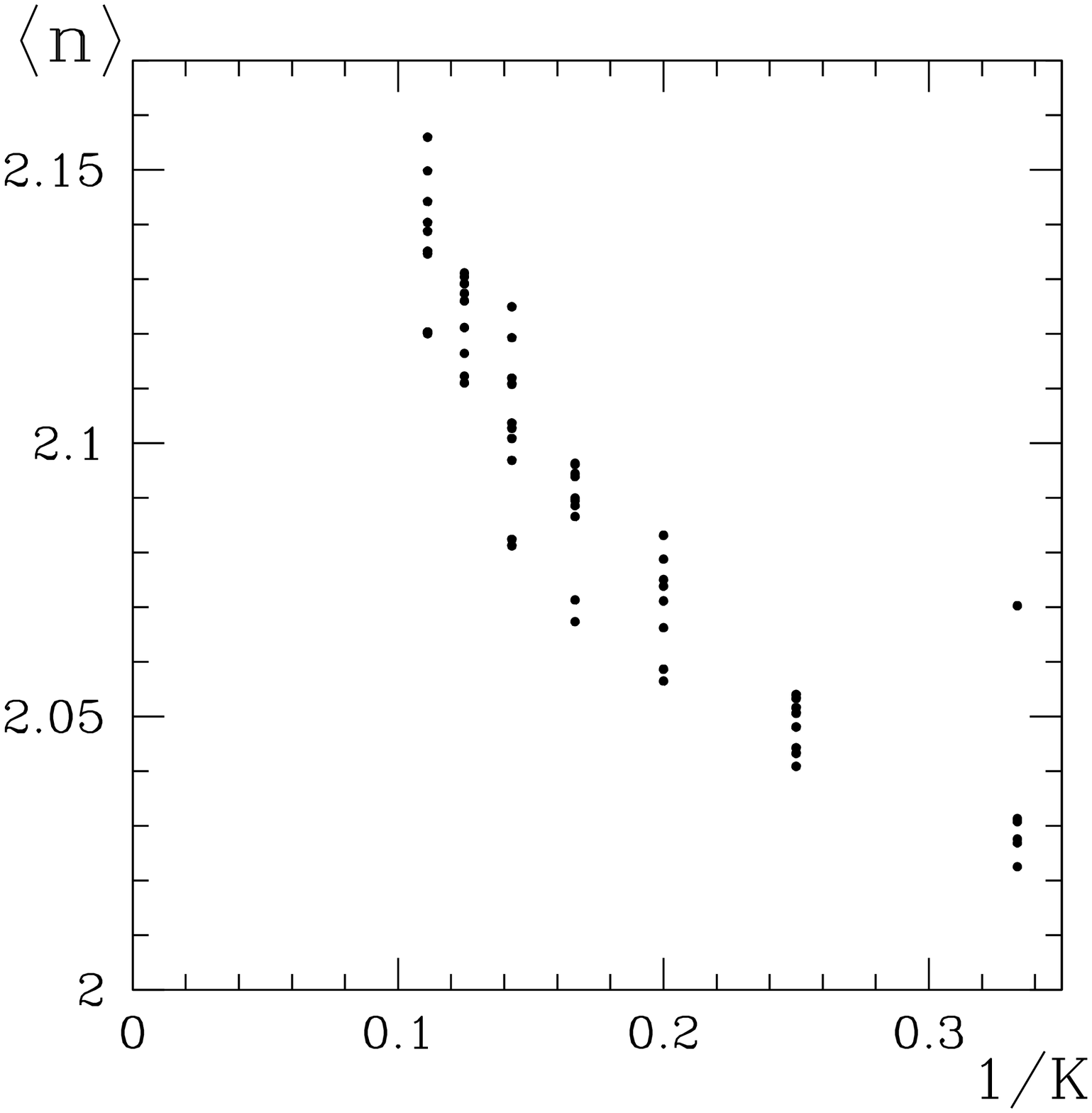,width=7.5cm,angle=0}&
\psfig{figure=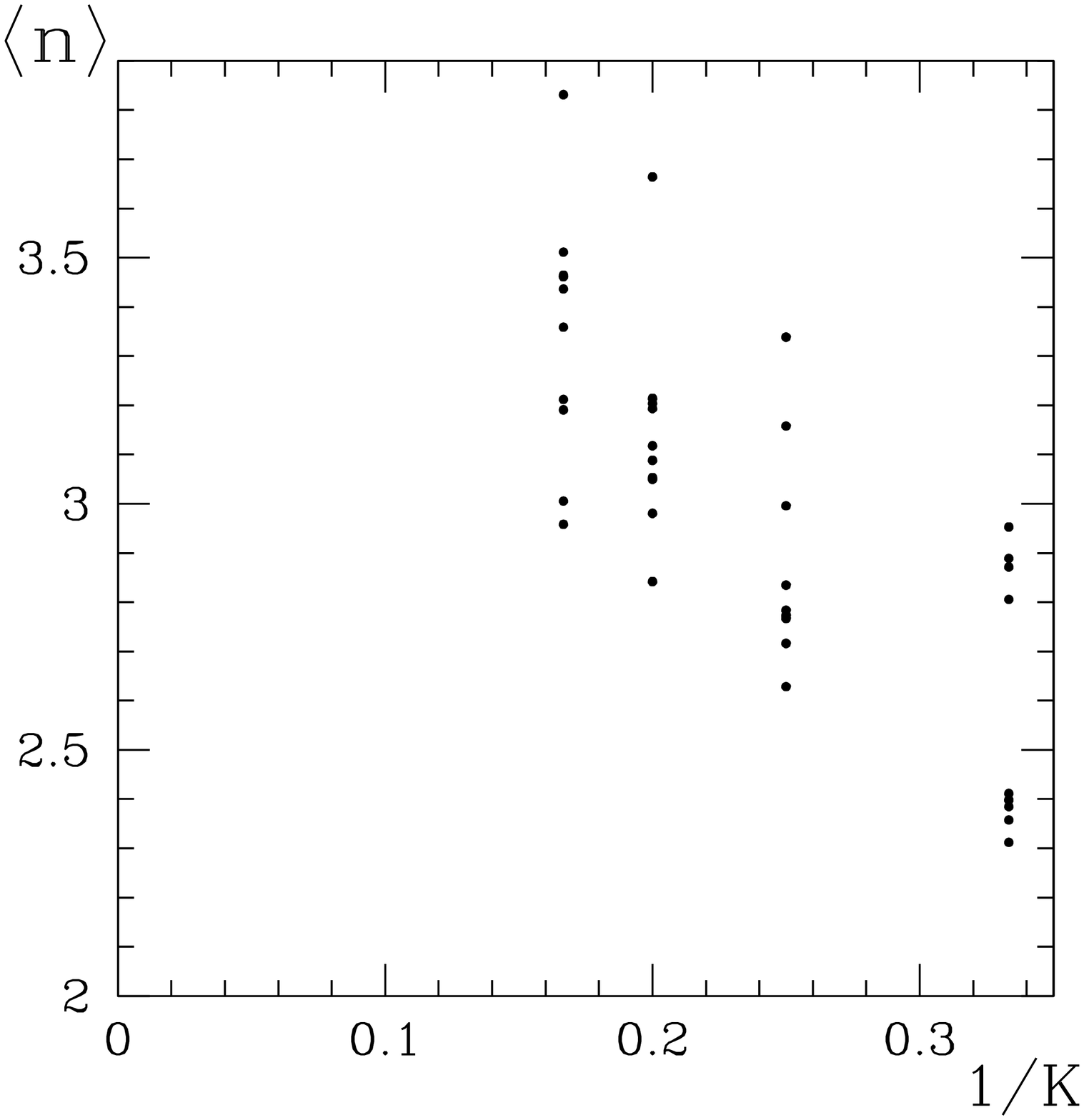,width=7.5cm,angle=0}\\
(a)&(b)
\end{tabular}
\caption{The average number of particles in the 10 lowest bound states  
vs.\ $1/K$ for (a) $g=0.1$ and (b) $g=0.5$ in units of
$\sqrt{4\pi^3/N_cL}$.  The Chern--Simons coupling is
$\kappa=2\pi/L$.\label{averageN} }
\end{figure}

We will now perform a detailed investigation of the strong-coupling 
spectrum at $g=0.5$.  At this stronger coupling, however, we are unable 
to make the additional approximation of cutting off the number of partons 
at 4 particles.  The reason for this is that as we
increase $K$ the average number of partons in a state 
also increases as we described earlier. We show
this behavior in Fig.~\ref{averageN}, and we see that at the 
largest values of $K$ the average number of particles is 
approaching 4. This tells us that, if we cut off the
number of particles at 4, we are missing a significant part of 
the wave function, and the approximation is starting to break down. 
We have checked this by looking at the spectrum with and without 
a 4-parton cutoff at $K=6$, and we find that we get differences as
large as $10\%$. In the region where the approximation remains valid,
however, we have a good approximation, and all the fits are linear. 
We can use these to extrapolate to the continuum. While we have always 
been able to make this linear fit in SDLCQ,  it remains an
assumption that linear fits remain valid outside of any region where 
we cannot calculate. Keeping 6 particles at resolution $K=6$ means 
that we are only able to go to a transverse resolution of $T=4$.

\begin{figure}
\begin{tabular}{cc}
\psfig{figure=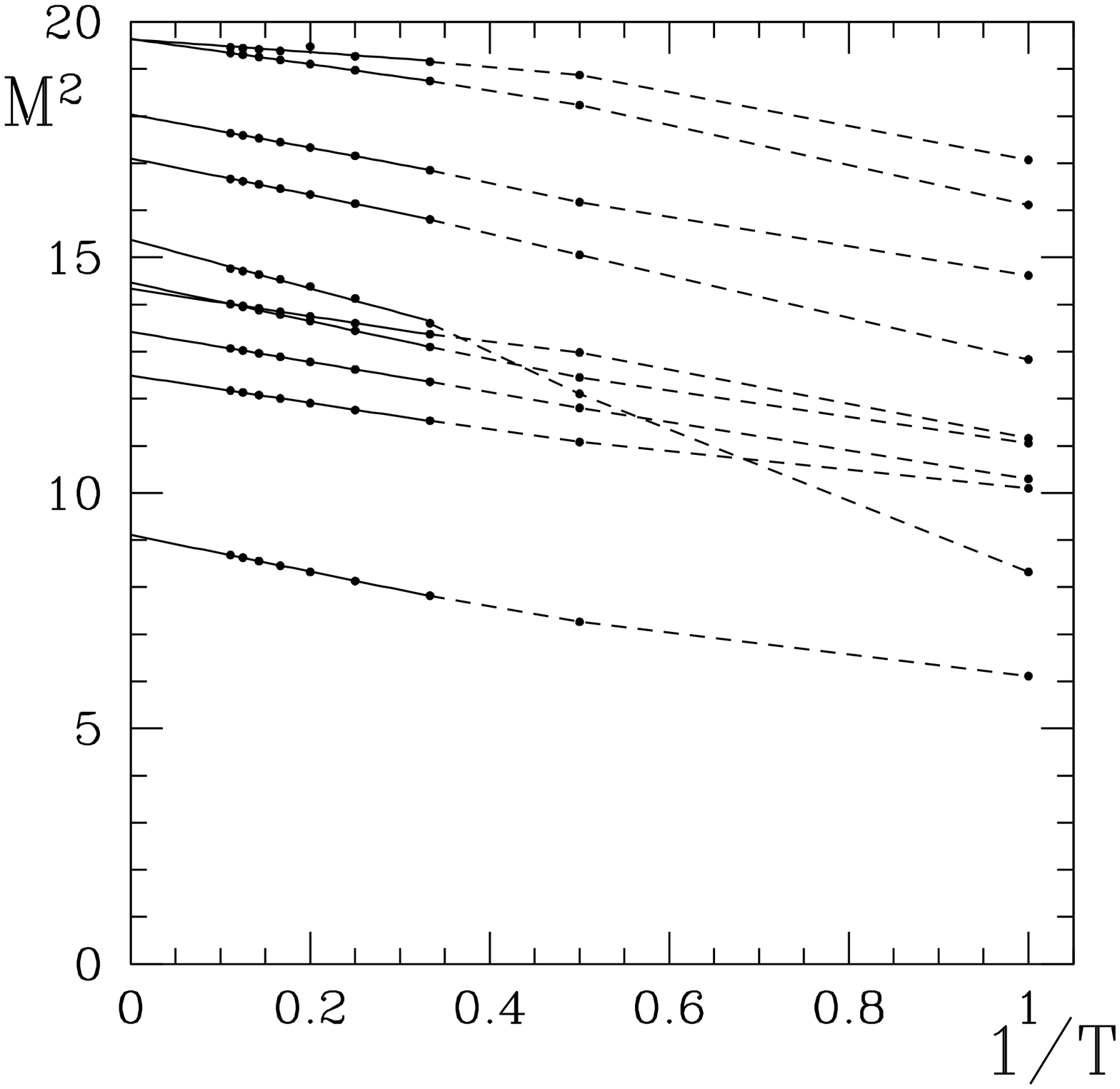,width=7.5cm,angle=0}&
\psfig{figure=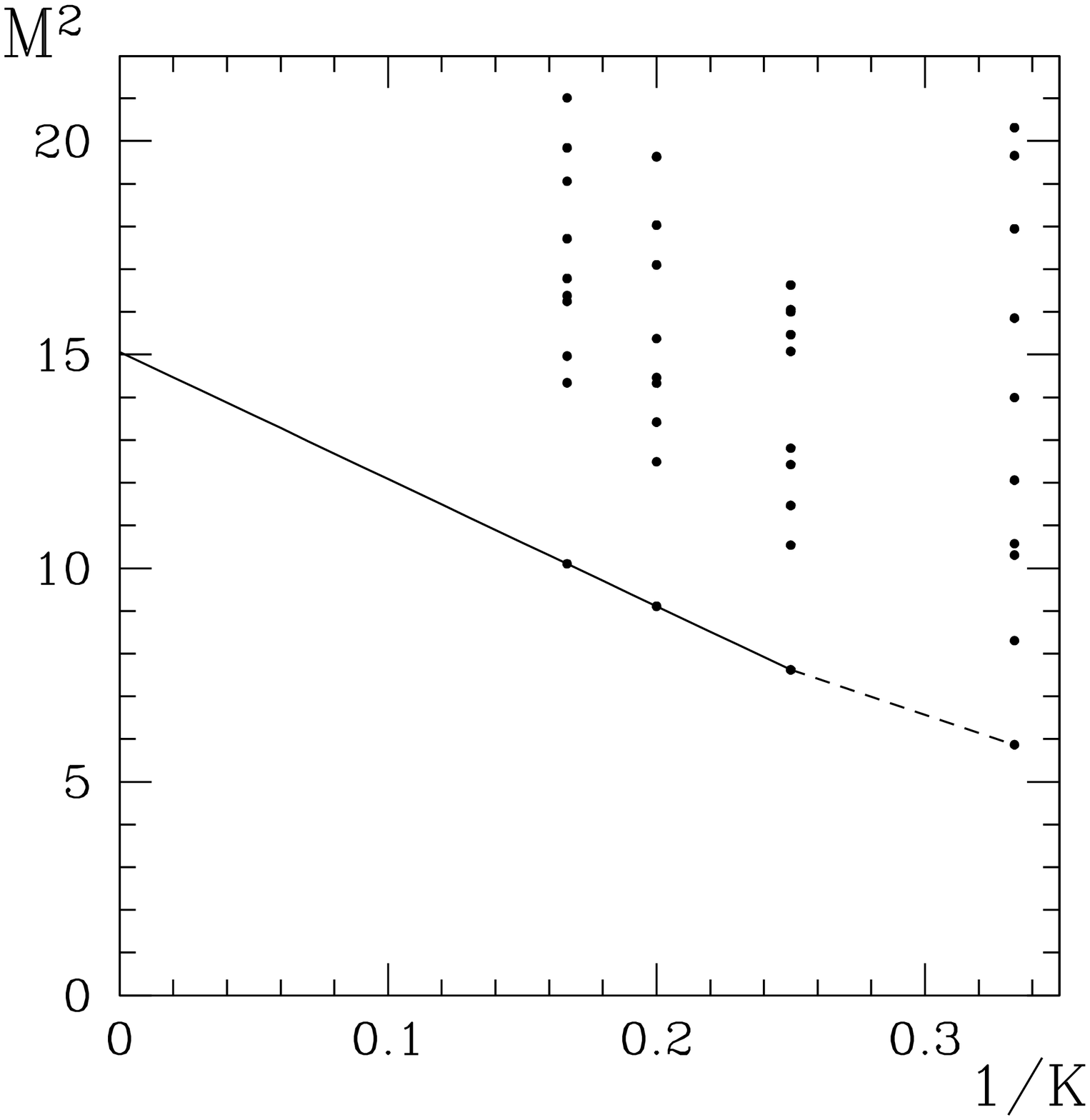,width=7.5cm,angle=0}\\
(a)&(b)
\end{tabular}
\caption{The spectrum as a function of (a) $1/T$, with a linear fit 
starting at $T=3$, and (b) the longitudinal resolution $K$, with a linear
fit for the lowest state.  The coupling values are $g=0.5\sqrt{4\pi^3/N_cL}$ 
and $\kappa=2\pi/L$. \label{g5}
} 
\end{figure}

At $g=0.5$ we repeat the standard analysis of the 10 lowest mass
states in each of the  sections. The spectrum as a function of the 
transverse cutoff $T$ is shown in Fig.~\ref{g5}, and we again see 
that the curves are well described by a
linear fit in $1/T$. The infinite transverse cutoff values of the 
masses are plotted as a function of $1/K$; we find good linear 
fits for the lowest state indicating that we again have good 
convergence. A linear fit for the higher
states appears less convincing and clearly would benefit 
from additional data. We have done the same analysis in the 
$Z_2=-1$ sector, and we will present some of the
results later in this section.

It is important at this point to
comment on a feature that appears in Fig.~\ref{Mvsg}.
\begin{table}
\centerline{
\begin{tabular}{|c||c|c|c|c|c|c|c|c|c|c|}\hline
$g$ & \multicolumn{10}{|c|}{$M^2$}\\\hline
$0.1$ & 4.36 & 4.85 & 4.93 & 7.41 & 7.47 & 8.35 & 8.36 & 9.34 & 9.43 &
9.48\\ 
\hline
$0.5$ & 8.93 & 13.14 & 13.88 & 14.92 & 15.10 & 15.62 & 15.82  & 17.38 &
18.35
&19.76\\ 
\hline
\end{tabular}}
\caption{\label{gap}
Mass squared in units of $4\pi^2/L^2$ of the lowest mass states
at $g=0.1$ and $g=0.5$, in units of $\sqrt{4\pi^3/N_c L}$ in the $Z_2=+1$
sector, 
showing the development of a gap above the lowest
state. Results in both cases are for resolutions $K=6$ and $T=4$ and for
Chern--Simons coupling $\kappa=2\pi/L$.}
\end{table}
We see a decoupling of the lowest state from the other states in 
the spectrum, whose masses grow more rapidly with coupling. In addition 
there is one low-mass state that appears to fall with increased 
coupling and become nearly degenerate with the lowest mass state. 
Calculations at larger values\footnote{For these increased
resolutions the time required for diagonalizations at many 
values of $g$ is prohibitive; therefore, we studied the $T$ dependence
at only two values of $g$.}  
of $T$ indicate that this behavior is a numerical artifact and
disappears when we increase $T$.  With this state removed 
the gap between the lowest state and the remaining states is even more 
apparent. In Table~\ref{gap} we present the
masses of the 10 lowest mass bound state that we obtained at $K=6$, $T=4$
for $g=0.1$ and $g=0.5$. We see that at weak coupling the states are all
closely spaced while at strong coupling a gap appears. From Fig.~\ref{g5}(b) 
it appears that gap is in fact increasing with $K$ so that in the 
continuum limit the effect is even larger.

The low-mass state separating from the rest of the spectrum is 
related to an effect that we saw in SYM-CS theory in 1+1 
dimensions~\cite{SYMCS1+1,BPS1+1}. There 
most of the spectrum behaves like $g^2$. There was, however, a special 
set of states, which we called approximate BPS states, that were nearly 
constant in mass as a function of the coupling. They are a reflection 
of the exactly massless BPS states in SYM theory in
1+1 dimensions. In the limit of very large coupling we found that 
the spectrum of these approximate BPS states was proportional to 
the average number of particles in the bound state. In 1+1 dimensions 
the average number of particles  was itself independent of the coupling.

In the (2+1)-dimensional SYM theory without a CS term 
there are again exactly massless BPS 
states~\cite{hpt2001,hpt2001b}. The average number of 
particles in these and other states grows with the coupling.  This 
increase in the average number of particles causes the approximate 
BPS states to increase in mass as we increase the coupling.
This increase is slower, however, than the $g^2$ behavior we see 
in the other states.

In the $Z_2=-1$ sector the lowest energy state is a three-parton state.
Here again we see the same approximate BPS phenomena that we saw in 
the $Z_2=1$ sector.  In Table~\ref{gap2} we see that at small coupling 
all the masses of the states are closely spaced
but at strong Yang--Mills coupling a gap develops between the lowest 
and the remaining states. There is again a state with anomalously light 
mass at strong coupling.  The effect here is not
as dramatic because the states are more massive and the gap
is a smaller relative to the masses.

Presumably there are higher mass approximate BPS states in this theory, 
but at coupling of the order $g=0.5$ they are intermixed with normal 
states, and it is difficult to study
them. At large enough coupling of course we expect them to separate, but as
discussed we have problems going to higher couplings.
\begin{table}
\centerline{
\begin{tabular}{|c||c|c|c|c|c|c|c|c|c|c|}\hline
$g$ & \multicolumn{10}{|c|}{$M^2$}\\\hline
$0.1$ & 9.61 & 11.29 & 11.40 & 11.50 & 11.61 & 13.70 & 13.77 & 15.57 & 15.58
& 
15.60\\ 
\hline
$0.5$ & 15.71 & 19.24 & 19.78 & 21.01 & 21.60 & 22.16 & 22.82  & 23.53 &
23.93
&24.31\\ 
\hline
\end{tabular}}
\caption{\label{gap2}
Same as Table~\ref{gap}, but for $Z_2=-1$.}
\end{table}
A more complete discussion of the approximate BPS states in 
(2+1)-dimensional SYM-CS theory can be found in~\cite{BPS2+1}.

\subsection{Structure Functions} \label{structure}
In this section we will discuss the structure functions as functions 
of both the longitudinal and transverse momentum. We will look at both 
$g=0.1$ and $g=0.5$. The structure function is basically the probability 
that a constituent of type A has a longitudinal momentum fraction $x=k^+/P^+$ 
and transverse momentum $k^\perp$.  It is
given as follows in terms of the light-cone wave function $\psi$:
\begin{eqnarray}
g_A(x,k^\perp)&=&\sum_q\int_0^1 dx_1\cdots dx_q
\int_{-\infty}^{\infty}
dk^\perp_1\cdots dk^\perp_q
\delta\left(\sum_{i=1}^q x_i-1\right)
\delta\left(\sum_{j=1}^q k^{\perp}_j\right)\nonumber \\
&&\qquad\times
\sum_{l=1}^q \delta(x_l-x)\delta(k^\perp_l-k^\perp)\delta^A_{A_l}
|\psi(x_1,k^\perp_1;\ldots x_q,k^\perp_q)|^2\,,
\end{eqnarray}
where $A_l$ is the type for the $l$-th constituent.

\begin{figure}
\begin{tabular}{ccc}
\psfig{figure=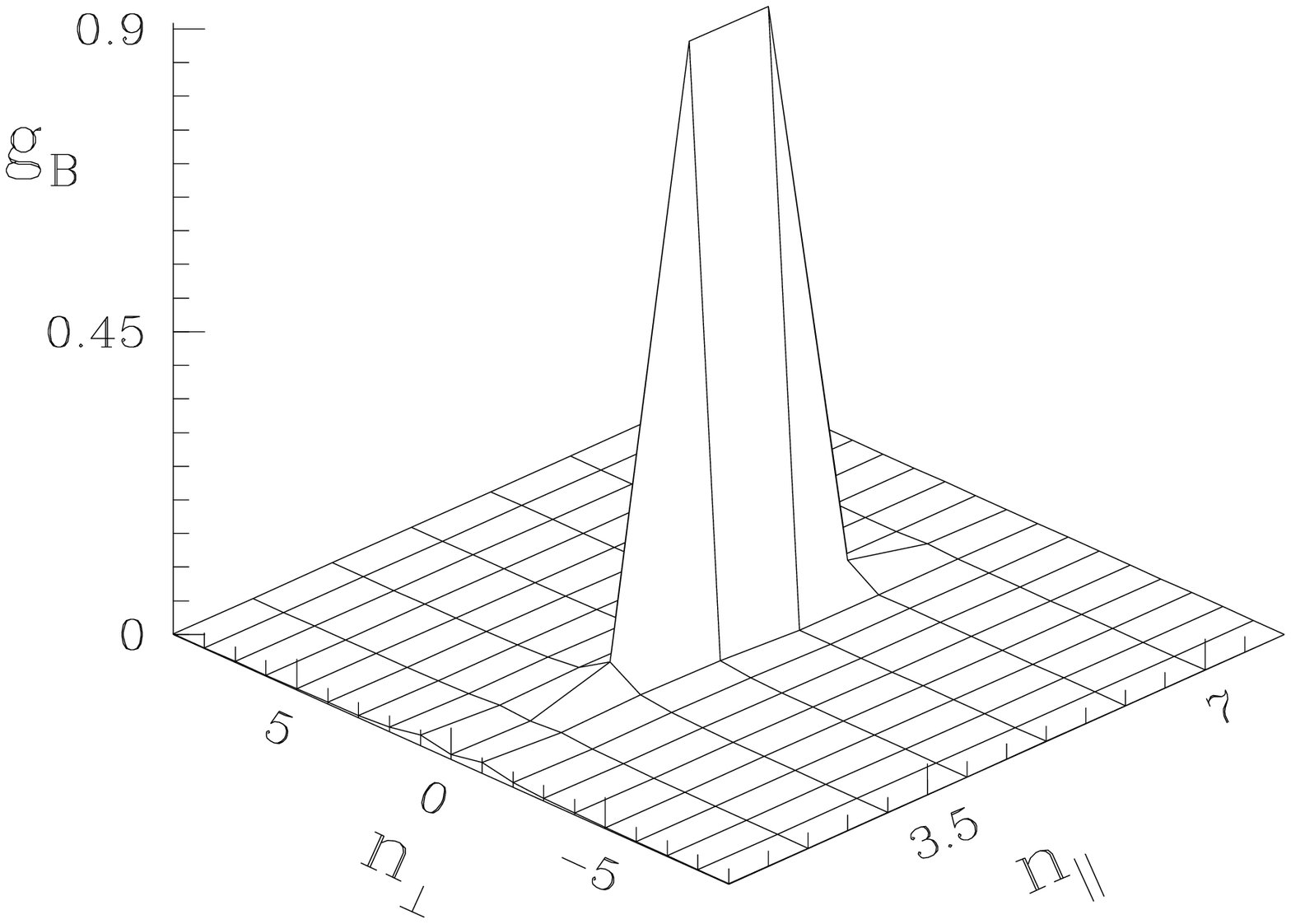,width=5cm,angle=0}&
\psfig{figure=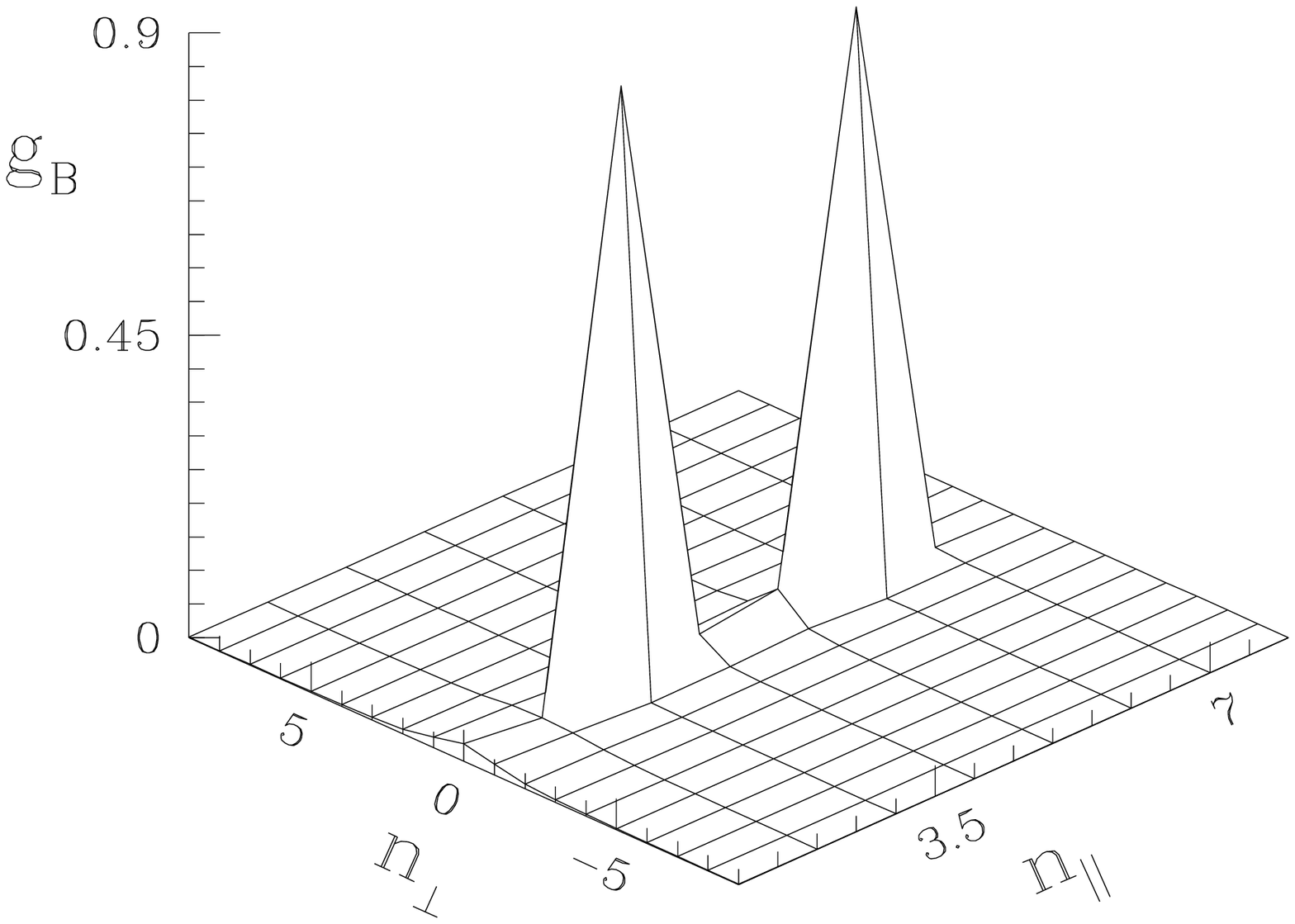,width=5cm,angle=0}&
\psfig{figure=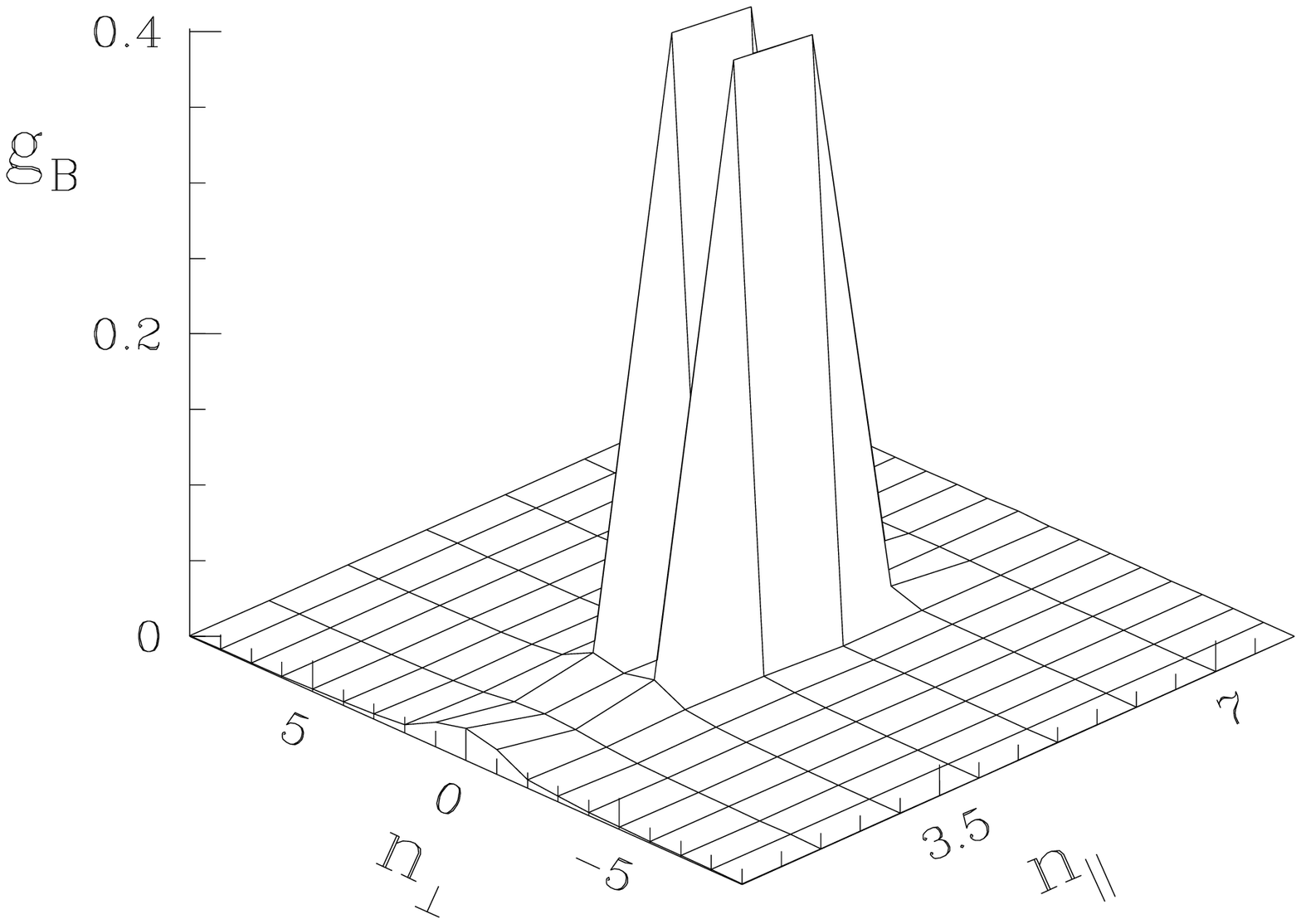,width=5cm,angle=0}\\
(a)&(b)&(c)
\end{tabular}
\caption{Bosonic structure functions of the lowest energy state
with (a) and (b) primarily two partons, 
and (c) two KK partons.  The symmetry sector is $Z_2=+1$,
and the couplings are $g=0.1\sqrt{4\pi^3/N_cL}$ and $\kappa=2\pi/L$.
The resolutions are $K=9$ and $T=9$.
\label{strg01}
}
\end{figure}

It is instructive to consider first the structure function for a few 
states at weak coupling. In Fig.~\ref{strg01} we see the structure 
function of two two-particle bound states and a
bound state composed of KK particles.  Fig.~\ref{strg01}(a) is the lowest
mass state in the $Z_2=1$ sector has  primarily two particles, one with
longitudinal momentum $n_{||}=4$ and one with longitudinal momentum 
$n_{||}=5$, and no units of transverse momentum.  Fig.~\ref{strg01}(b) is 
another two-parton bound state with clearly separated peaks.
The partons have $n_{||}=3$ and $n_{||}=6$ units 
of longitudinal momentum  and no units of 
transverse momentum. The state in Fig.~\ref{strg01}(c) is very similar 
to the bound state shown in Fig.~\ref{strg01}(a) except that it is a bound state 
of particles that  have $n_\perp=1$ and $n_\perp=-1$ units of transverse
momentum and is therefore an example of a state that is viewed in 
two dimensions as a bound state of KK particles. These structure 
functions are very simple because
at weak coupling these bound states are composed of nearly free partons.

As we move to strong coupling, the structure functions change 
significantly. An overall effect which we see in all the 
strong-coupling structure functions is the stringiness of the states, 
i.e.~states have many partons.  Therefore the average momentum per 
parton decreases, and we see the peaks in
the structure function move to lower $n_{||}$. Within the 
range of parameters we have selected here, the interaction energy 
at $g=0.5$ is as large as the KK excitation
energy. Therefore we will see mixing between the KK modes and the low-energy
modes.

\begin{figure}
\begin{tabular}{ccc}
\psfig{figure=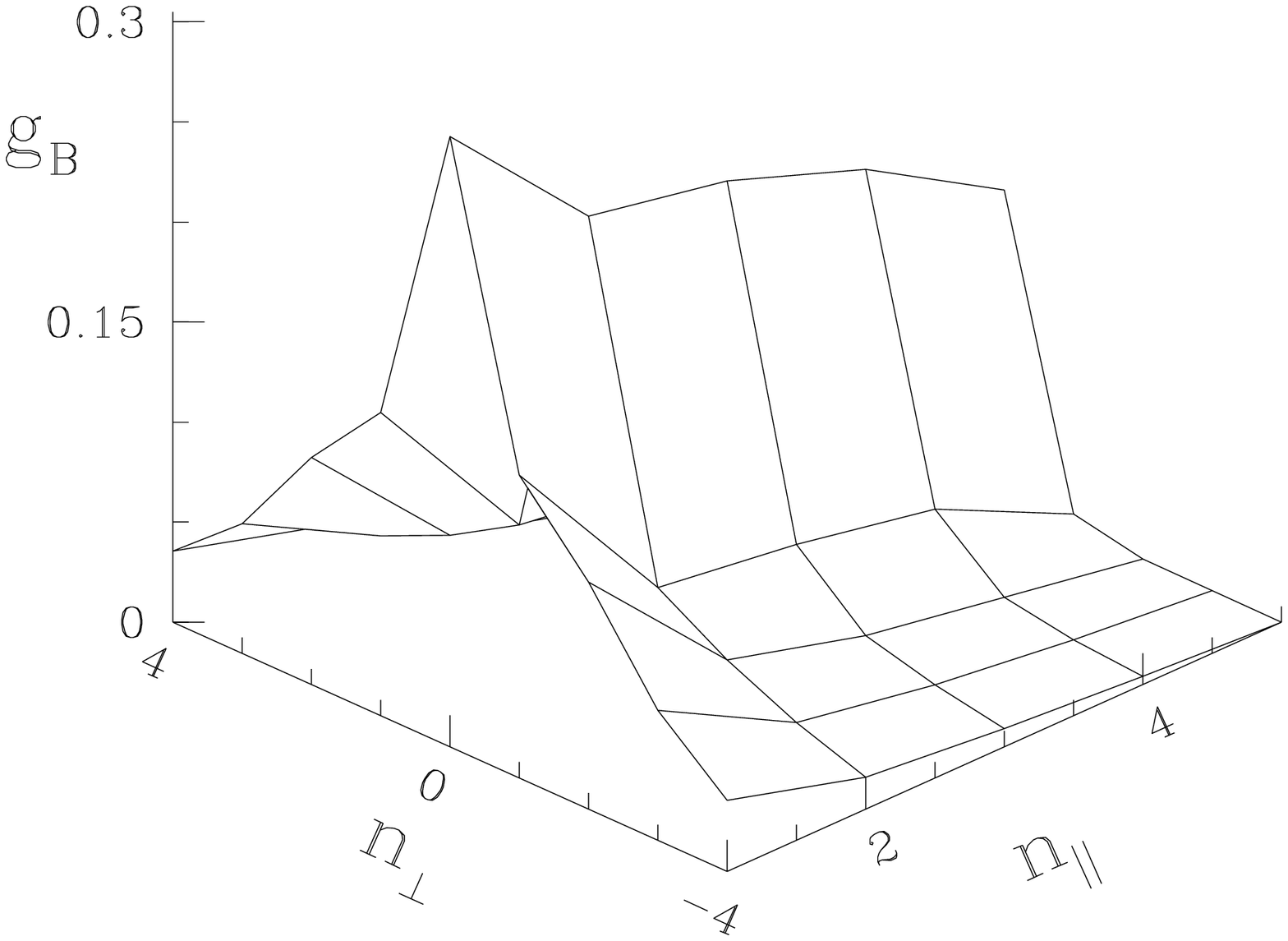,width=5cm,angle=0}&
\psfig{figure=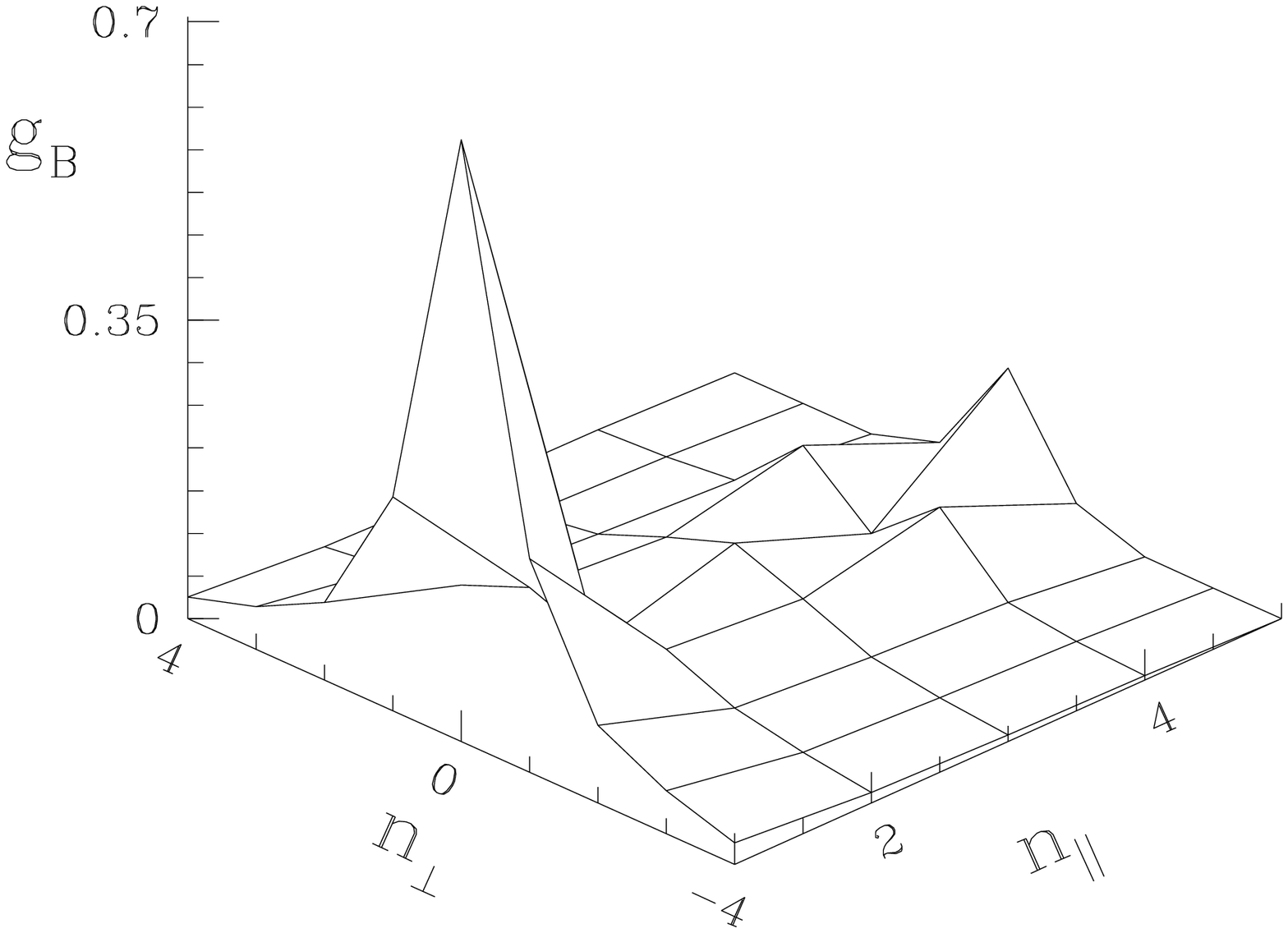,width=5cm,angle=0}&
\psfig{figure=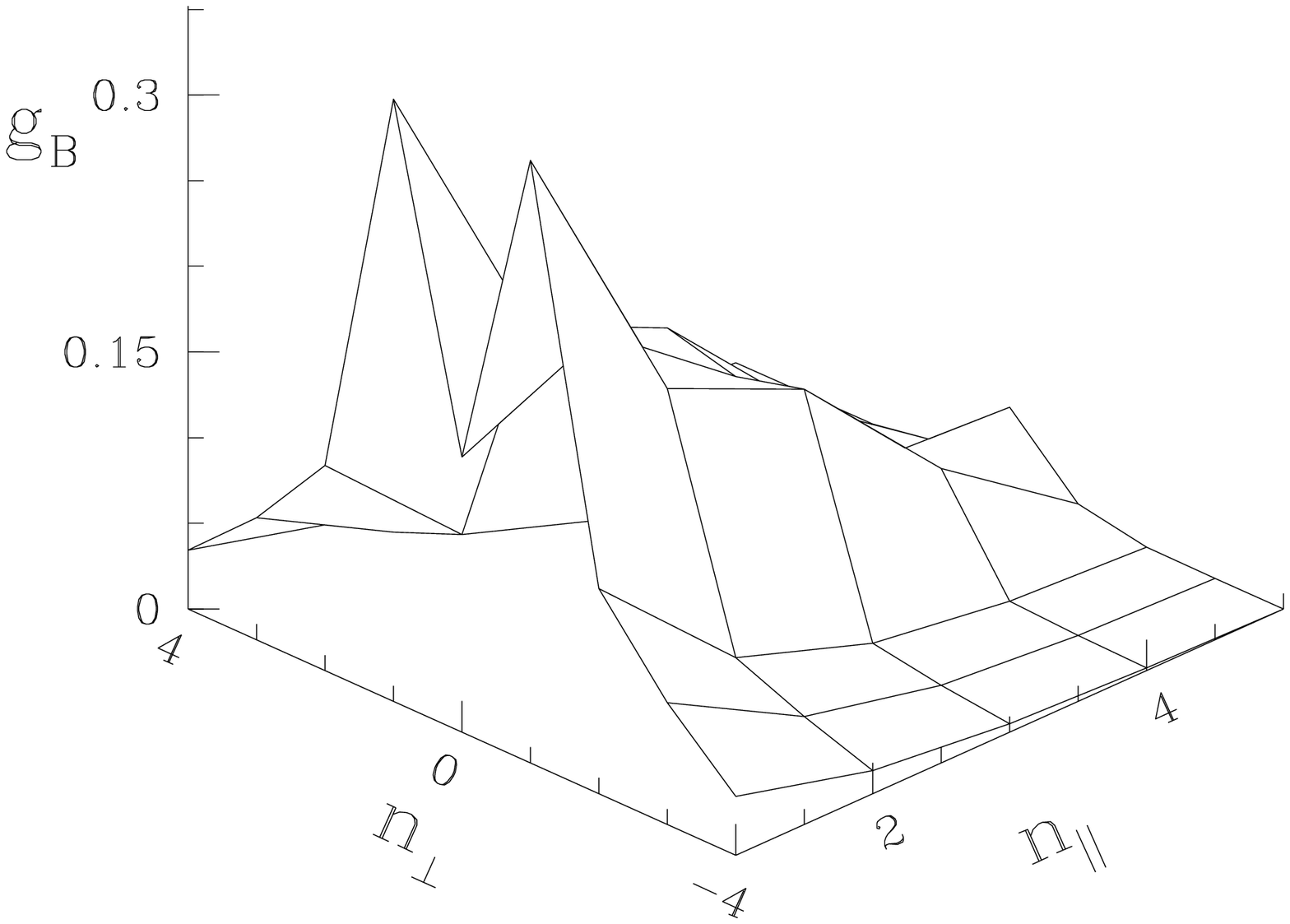,width=5cm,angle=0}\\
(a)&(b)&(c)
\end{tabular}
\caption{Same as Fig.~\ref{strg01}, but for $g=0.5\sqrt{4\pi^3/N_cL}$
and resolutions $K=6$ and $T=4$.
\label{strg02}
} 
\end{figure}

In Fig.~\ref{strg02}(a) we see the structure function of the lowest 
energy two-parton state at $g=0.5$. As we discussed earlier this is an 
anomalously light state and is a reflection of the BPS states in the pure 
SYM theory.  Since it is so light it does not mix with the KK modes, and 
it is very flat in $n_{||}$. This flatness seems to be unique to this state. 
The state in Fig.~\ref{strg02}(b) appears to be  at least partially a 
three-parton state. The shape of its structure function is consistent
with having a component with  $n_{||}=4$ and $n_\perp\pm 1$,
$n_{||}=1$ and $n_\perp\pm 1$, and $n_{||}=1$ and $n_\perp=0$. There also
appears to be a two-parton component with $n_\perp=0$ for both partons
and $n_{||}=5$  and $n_{||}=1$. This is therefore an example of 
mixing between the KK sector and the low-energy sector as well as a mixing 
between the two-parton and three-parton sectors.  Finally, in 
Fig.~\ref{strg02}(c) we have a state that is dominantly a KK state. It has 
the lowest energy for such a state and is therefore most likely the evolution 
of the pure KK state we saw at $g=0.1$. We note again that there is mixing 
with the low-energy sector, since there is a significant probability 
of finding a parton with one unit of $n_{||}$ and no transverse momentum.

It is clear from this study that
a search for a signal of extra dimensions in the form of KK
bound states may be frustrated by strong mixing with 
ordinary states, a situation reminiscent of the long-standing
hunt for glue balls. 
One hopes that the energy scales are well separated, as in the weak-coupling 
example, so that these bound states of KK particles can be identified from
the mass spectrum alone.
 
%

\begin{figure}
\begin{tabular}{ccc}
\psfig{figure=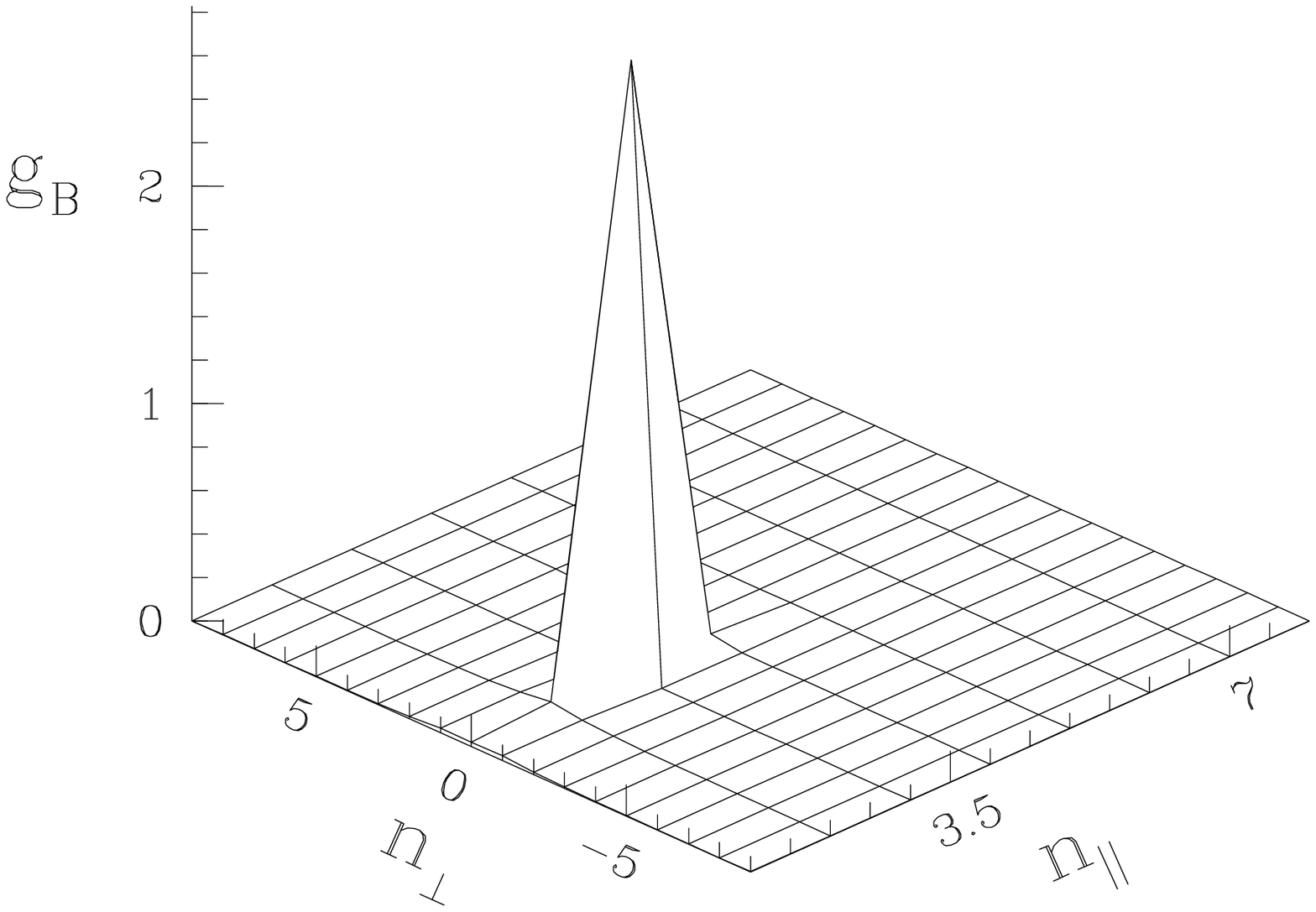,width=5cm,angle=0}&
\psfig{figure=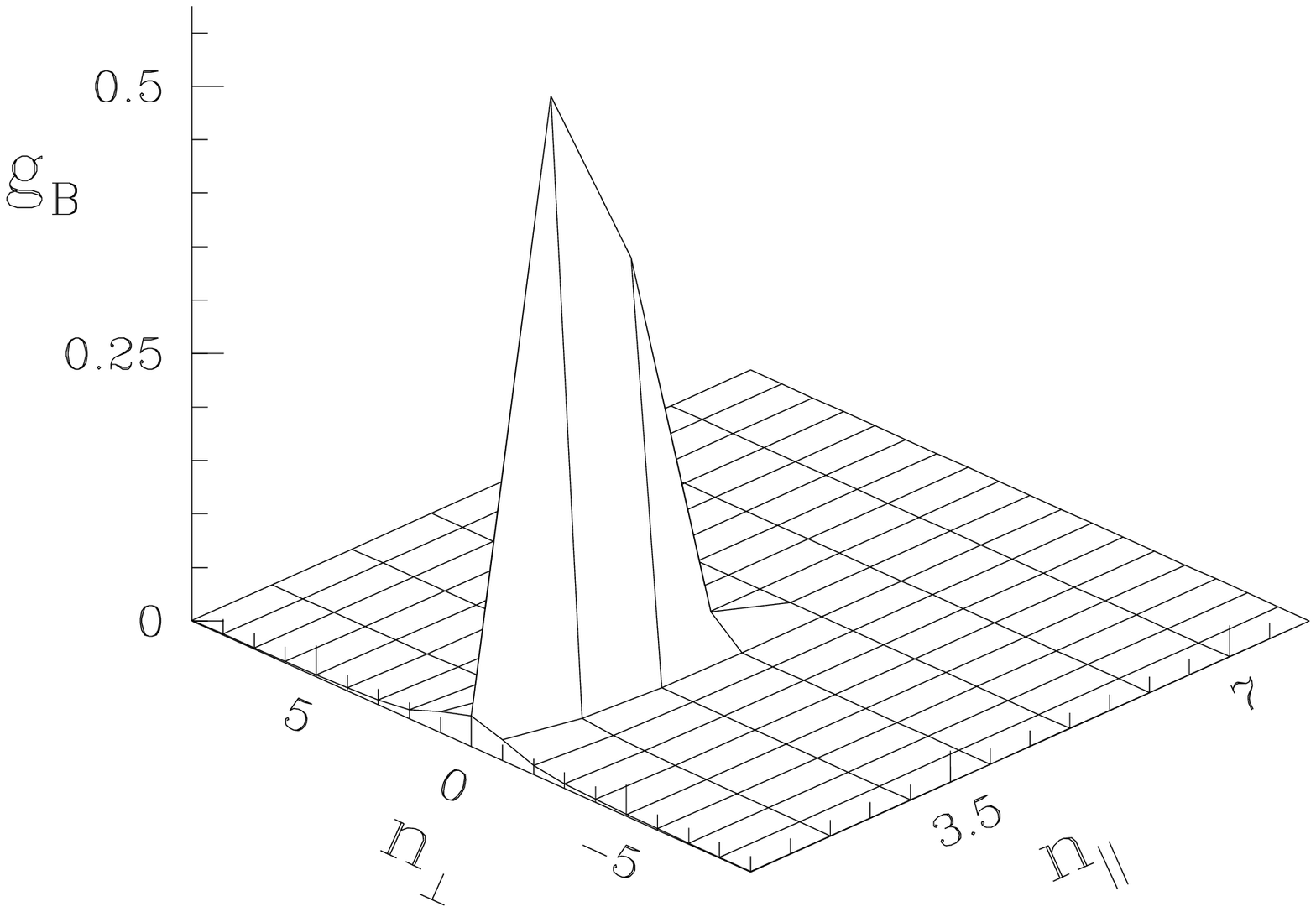,width=5cm,angle=0}&
\psfig{figure=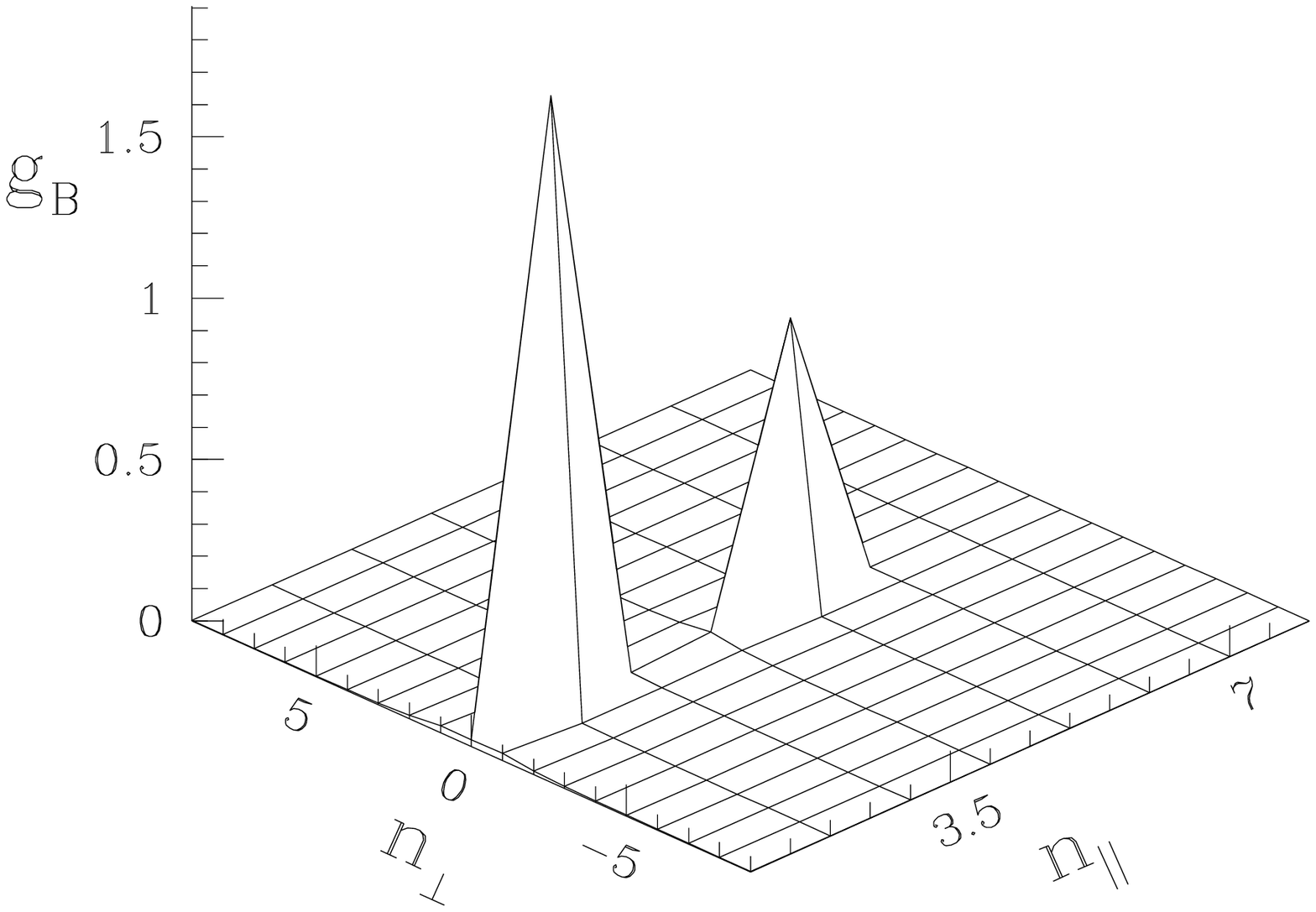,width=5cm,angle=0}\\
(a)&(b)&(c)
\end{tabular}
\caption{Structure functions for the lowest energy states
with primarily three partons.  The masses squared are
(a) 9.8, (b) 10.54, and (c) 11.61, in units of $4\pi^2/L^2$.
The symmetry sector is $Z_2=-1$, and the couplings 
are $g=0.1\sqrt{4\pi^3/N_cL}$ and $\kappa=2\pi/L$.
The resolutions are $K=9$ and $T=9$.
\label{strg03}
} 
\end{figure}

\begin{figure}
\begin{tabular}{ccc}
\psfig{figure=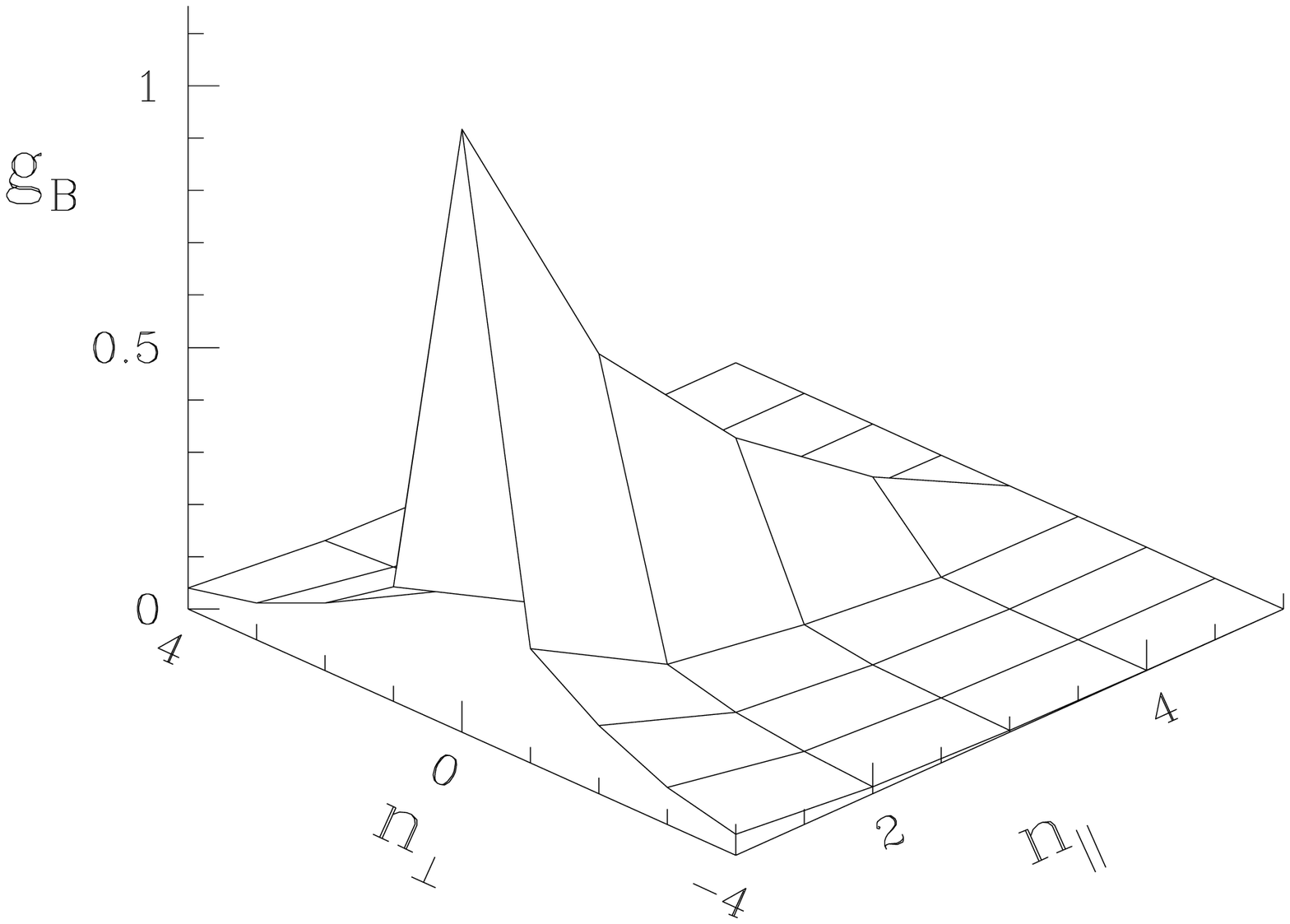,width=5cm,angle=0}&
\psfig{figure=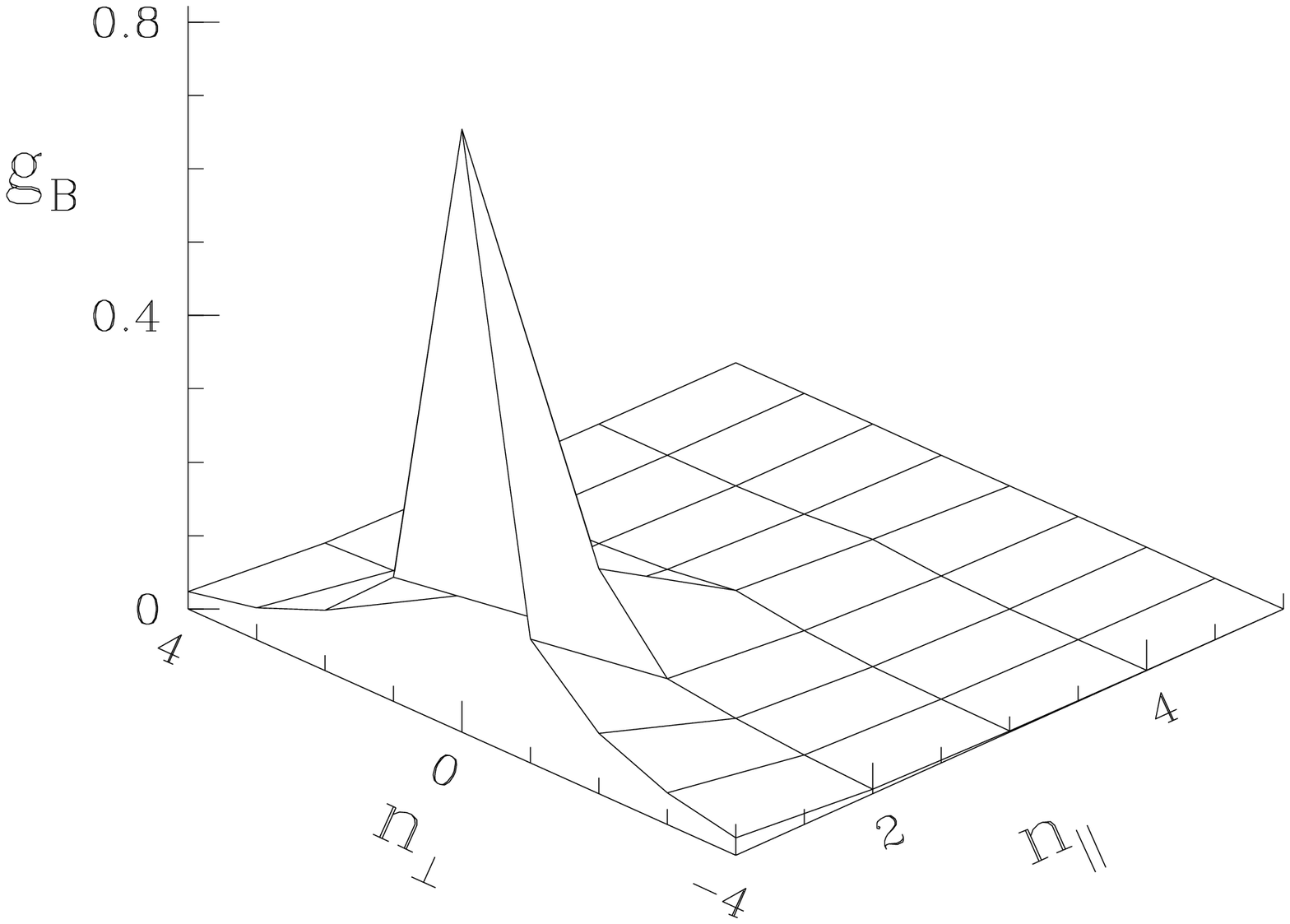,width=5cm,angle=0}&
\psfig{figure=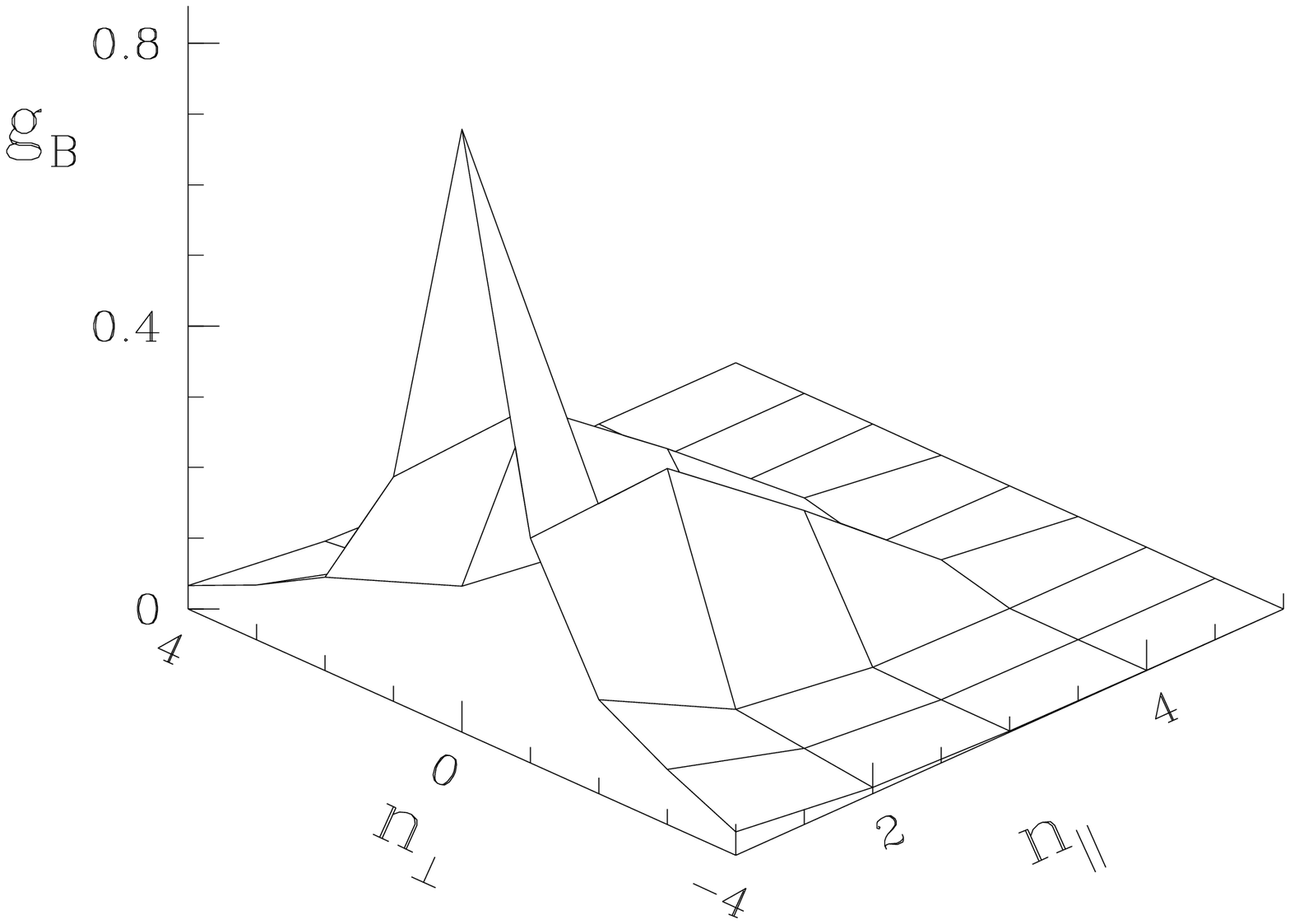,width=5cm,angle=0}\\
(a)&(b)&(c)
\end{tabular}
\caption{Same as Fig.~\ref{strg03}, but for the corresponding
states at Yang--Mills coupling $g=0.5\sqrt{4\pi^3/N_cL}$
and resolutions $K=6$ and $T=4$.  The mass values are
(a) 15.71, (b) 19.25, and (c) 22.60.
\label{strg04}
} 
\end{figure}

In the $Z_2=-1$ bosonic sector the lowest states are three-particle states at 
weak coupling which broaden to  stringy states as the coupling is increased, 
just as in the $Z_2=+1$ sector. There are  several interesting structure 
functions in this sector,  both at weak coupling and at strong coupling. They 
are shown in Figs.~\ref{strg03} and \ref{strg04}.  The lowest energy state at 
$g=0.1$ is a classic pure three-parton state.  The state at $M^2=10.544$ is a 
three-parton state with bosons at $n_{||}=2$ and 3 as well as a fermionic 
component that does not show up in this structure function.  The state at
$M^2=11.606$ is interesting because it is clearly a  pure state with two 
partons with 2 units of $n_{||}$ and one with $n_{||}=5$ units.

The three-parton state in Fig.~\ref{strg04}(a)
at $g=0.1$ is an approximate BPS state
and therefore has an anomalously light mass relative to the other states at
$g=0.5$. The structure function has the characteristic long ridge in
$n_{||}$ that we saw in the $Z_2=1$ sector. 
The peak of structure function in Fig.~\ref{strg04}(b) has moved to smaller $n_{||}$
which is a characteristic of these supersymmetric bound states at strong coupling. 
We have included Fig.~\ref{strg04}(c) because it has a high enough  mass that it
mixes with the KK modes in this sector. 

\section{Summary} 
\label{sec:summary}

We have considered SYM-CS theory in 2+1 dimensions and presented
numerical solutions for the spectrum and structure functions, as well other
properties of the bound states. There are several motivations for this
calculation.  Previously we found two very interesting
features in the solutions of SYM in 2+1 dimensions~\cite{BPS2+1}. 
The states become very stringy as we increase the Yang--Mills coupling,
in the sense that the average number of partons in all the bound states 
grows rapidly.  The question is whether this property will change
significantly when we add a mass for the partons. 
Therefore the technical question that we are attempting 
to answer here is how these properties of the solutions of SYM 
theory in 2+1 dimensions change when we add a CS term, which effectively 
gives a mass to partons of the theory without
breaking the supersymmetry.   From a physics point of view we are looking
for new phenomena that might appear, particularly phenomena that are too 
complicated to be investigated analytically.

At weak Yang--Mills coupling we find a spectrum that is in many ways 
very similar to the free spectrum. Namely, we find  bound states at 
weak coupling which are very close in mass
to the discrete approximation to the multi-particle continuum but with the
squares of the masses shifted by approximately 
%
%
$\Delta M^2{=}2\kappa g \sqrt{N_c}$.  We find a
full spectrum of KK states above the low-energy states.  Since KK
states are of considerable interest within the context of extra dimensions, we
investigated the relation between the low-energy states and the KK states.
This could be considered as a toy model of universal extra dimensions in 
the gauge sector, with the extra twist that the toy model is fully 
supersymmetric. At weak coupling the KK states are well separated 
from the low-energy states, and their spectrum is
easily understood in terms of the low-energy spectrum. As we go to larger
coupling we find extensive level crossing and beyond this what we call 
the strongly coupled region. In the numerical
calculation we have taken the interaction energy to be of the same 
scale as a unit of KK mass.  We find that the KK states and the 
low-energy states mix. We looked at the structure functions of these states 
and found that the structure functions give a
clear picture of the mixing. When viewed from a lower number of dimensions,
there does not appear to be a simple way to disentangle the KK states 
from the low-energy states.

We studied the strong-coupling region in considerable detail and find
another interesting feature. The BPS states of the underlying SYM theory
become states with anomalously low mass. These states are 
the (2+1)-dimensional versions of the approximate BPS states that we saw 
in the dimensionally reduced theory~\cite{SYMCS1+1,BPS1+1}. 
We looked at the lowest such states 
in both $Z_2$ sectors. It seems entirely possible that this could be a 
novel mechanism for generating states at a new scale significantly below 
the fundamental scale of a theory. The structure functions of these states 
appear to be unique in the sense that they are
much flatter in $n_{||}$ than those of the other bound states of the theory.

There are a number of interesting research directions from here. The CS 
term is clearly an efficient way to introduce a mass without breaking
supersymmetry, and it would be interesting to investigate coupling this 
gauge sector to fundamental matter.

\section*{Acknowledgments}

This work was supported in part by the U.S. Department of Energy
and by grants of computing time by the Minnesota Supercomputing Institute. 
One of us (S.P.) would like to acknowledge the Aspen Center for Physics 
where part of this work was completed.

\end{document}